\documentclass{aa}  

\usepackage{graphicx}
\usepackage{txfonts}
\usepackage{amsmath}
\usepackage{amssymb}
\usepackage{natbib}
\usepackage{lipsum}
\usepackage{hyperref}

\usepackage{subcaption}         
\usepackage{lscape}             
\usepackage{placeins}           
                                
\usepackage{hyperref}

\graphicspath{{./figures/}}
\bibpunct{(}{)}{;}{a}{}{,}
\newlength{\minuslength}
\begin{document}

   \title{A 3D physico-chemical model of a pre-stellar core}
\subtitle{II. Dynamic chemical evolution in a pre-stellar core model using tracer particles}


   \author{S. S. Jensen\inst{1}\thanks{\email{sigurdsj@mpe.mpg.de}} 
   \and S. Spezzano\inst{1}
   \and P. Caselli\inst{1}
   \and T. Grassi\inst{1}
   \and O. Sipil{\"a}\inst{1}
   \and T. Haugb\o lle\inst{2}
   }

   \institute{Max-Planck-Institut f{\"u}r extraterrestrische Physik, Giessenbachstrasse 1, D-85748 Garching, Germany
   \and 
   Niels Bohr Institute, University of Copenhagen, Jagtvej 155A, DK-2200 Copenhagen, Denmark}

   \date{Draft date: \today}

 
  \abstract
   {Pre-stellar cores mark the earliest phase of star formation. By characterizing their physical and chemical structure, we can establish the initial conditions for star and planet formation and assess how closely the chemical composition of these cores is connected to later evolutionary stages.}
   {We explore the differences between static and dynamically evolving physico-chemical models of pre-stellar cores. The results are compared with observations of the pre-stellar core L1544 to estimate how well 3D physico-chemical models can reproduce the chemistry at this evolutionary stage.}
   {A 3D magnetohydrodynamic model of a pre-stellar core embedded in a dynamic star-forming cloud was post-processed using sequentially dust radiative transfer, a gas-grain chemical model, and a nonlocal thermodynamic equilibrium line-radiative transfer model. The chemical evolution was modeled along $\sim$20,000 tracer particle trajectories to capture the effect of a realistic dynamical evolution as the core formed. The emission morphology of CH$_3$OH and $c$-C$_3$H$_2$ and the intensities of CH$_3$OH, $c$-C$_3$H$_2$, CS, SO, HCN, HCO$^{+}$, and N$_2$H$^{+}$ were compared with observations of L1544. We compared initial elemental abundances with and without depletion of heavier elements.}
   {Our results show a distinct difference in chemical morphology between the dynamical and static models. The dynamical model reproduces the observed spatial distribution of CH$_3$OH and $c$-C$_3$H$_2$ toward L1544, whereas the static model fails to reproduce this morphology. In contrast, when we compared modeled and observed intensities across a broad range of molecules, the static model agreed well with observations for L1544. The dynamical model systematically predicts lower abundances and modeled intensities for six of the seven species presented here. For sulfur-bearing species, the intensities agree better with observations when the initial abundances are not depleted in heavier elements.}
   {We reveal distinct differences between dynamical and static physico-chemical models. The static model predicts higher abundances and intensities for the majority of the molecules we studied than the dynamical model. This discrepancy may stem from the specific choices of initial conditions, which might limit the ability of the dynamical models to fully capture the physical and chemical history. The intensities predicted by the static model are comparable to those observed toward L1544.}


\titlerunning{Dynamic chemical evolution in a pre-stellar core model using tracer particles.}
\authorrunning{S. S. Jensen et al.}

   \maketitle
   \nolinenumbers
%
\section{Introduction}\label{sec:introduction}

Pre-stellar cores are dense, gravitationally bound fragments of molecular clouds that are dynamically unstable and on their way to form stars or brown dwarfs. The study of the physical and chemical structure of pre-stellar cores therefore provides important constraints on the initial conditions of the star and planet formation process.

Previous studies of pre-stellar cores have revealed a rich chemistry, with a broad range of molecules detected, ranging from simple diatomic molecules and key ions (CO and HCO$^{+}$) to complex organic molecules (COMs) such as CH$_3$OH, CH$_3$OCH$_3$, and CH$_3$CHO \citep[e.g.,][]{2016ApJ...830L...6J}. These studies indicated that a significant fraction of the molecules available during planet formation may already be formed in the molecular cloud phase or during the pre-stellar core phase \citep{2012A&ARv..20...56C, 2021NatAs...5..684B, 2023ASPC..534..379C}. This is also supported by isotopic fractionation studies, which generally support a high degree of chemical inheritance \citep{2021A&A...650A.172J, 2021MNRAS.500.4901D, 2023ASPC..534.1075N, 2025NatAs.tmp..206L, 2025A&A...702A.127S}.

Recent studies have revealed a clear spatial segregation between different classes of molecules in pre-stellar cores \citep{2017A&A...606A..82S, 2020A&A...643A..60S}. 
The origin of the observed chemical structure is not yet understood. One hypothesis is that the observed segregation arises from the local cloud structure and environment. Star formation occurs across a wide range of different environments ranging from isolated Bok globules to dense star-forming regions with tens or hundreds of stars forming within parsec-scale regions (1-10 parsec). These diverse environmental conditions may affect the star formation process and the chemistry that evolves through a number of mechanisms: dynamical effects on the accretion process (density perturbations and differing timescales), and uneven irradiation through either structural differences in the core or cloud environment or because of local irradiation by nearby OB stars.
Another effect that might contribute to the chemical structure of pre-stellar cores is the dynamic nature of the accretion process itself. Recently, a growing number of accretion streamers have been detected toward pre-stellar cores \citep{2025A&A...696A.190C} and protostellar sources \citep{2020NatAs...4.1158P, 2024A&A...692A..55G}. Streamers may add fresh material with a distinct chemical composition, having undergone different physical and chemical histories \citep{2016ApJ...826...22K}.
To which extent these effects affect the chemical composition and structure of pre-stellar cores and the later stages of star formation is still an open question. In a recent study, \citet{2025A&A...699A.103G} applied a density-based clustering analysis to molecular emission maps toward three young cores. They reported a clear link between the H$_2$ column density $N$(H$_2$), the gradient in $N$(H$_2$), and the chemical morphology and associate this with environmental effects around the cores that shape the chemical morphology. 

To assess the effects of the local environment, models need to account for realistic cloud environments, which resemble the stellar nurseries we observe in the local galaxy. Furthermore, models need to account for a broad range of physics and chemistry, including magnetohydrodynamics (MHD), grain-surface chemistry, and photochemistry.
Because chemical inheritance likely plays a key role in the chemistry during star and planet formation, it is crucial to assess to which degree the local cloud environment might affect the early chemistry, and hence, ultimately the chemistry of the later planetary system.

While astrochemical models serve as an important tool for understanding observations, they regularly struggle to provide accurate predictions across multiple different species \citep[e.g.,][]{2015MNRAS.447.4004R, 2017ApJ...842...33V}. In these cases, the physical and chemical parameters may have to be fine-tuned to specific species or sub-categories of molecules to provide a good agreement with observed line intensities \citep[e.g.,][]{2013A&A...559A..53M, 2023A&A...676A..78G, 2025A&A...700A.141H}. These fundamental challenges may be a result of the many simplifications in astrochemical models. For example, rate equations for grain-surfaces reactions are crude approximations of the stochastic processes involved, and many parameters in these models are highly uncertain \citep{1998ApJ...495..309C, 2017SSRv..212....1C}. However, the apparent discrepancies in astrochemical models might also arise from limitations in the physical modeling. Often, 0D or 1D models are preferred to 3D models because they are simpler to develop and run and are often more straightforward to interpret \citep[e.g.,][]{2013ApJ...765...60G, 2025ApJ...990..163B}. 
It is essential to identify the primary source of limitations in current astrochemical models, regardless of whether they stem from the choice of physical model or from challenges such as poorly constrained chemical parameters and oversimplifications in chemical modeling. This understanding will guide the prioritization of the most critical problems to solve to improve the model.

We present a 3D MHD simulation of a pre-stellar core formed in a larger molecular cloud environment. The physical simulation is post-processed to compute the local temperature, radiation field, and the cosmic-ray ionization rates self-consistently around the core. The chemical evolution is then computed along tracer particle trajectories that accrete onto the core. This allows us to model the dynamical evolution of the matter that is accreted into the pre-stellar core in a realistic cloud environment. We compare the model with a previous static model of the same 3D core and to the numerous observations of the pre-stellar core L1544, located in the Taurus molecular cloud.

The paper is organized as follows. In Sect. 2 we introduce the model. In Sect. 3 we present the results of the model and compare them with the previous static model of the well-studied pre-stellar core L1544 and with observations of the same core. In Sect. 4 we discuss the implications of this work for future models, the comparison with L1544, and the next steps. In Sect. 5 we summarize our results.


\section{Model description} \label{sec:methods}

The overall framework of the model we used was previously introduced for the static model in \citet{2023A&A...675A..34J}. In this section, we provide a brief overview of the central elements of the model and a more in-depth description of the changes implemented in this work.

\subsection{Physical model}
\subsubsection{Molecular cloud simulation}
\begin{figure}[ht]
\resizebox{\hsize}{!}
        {\includegraphics{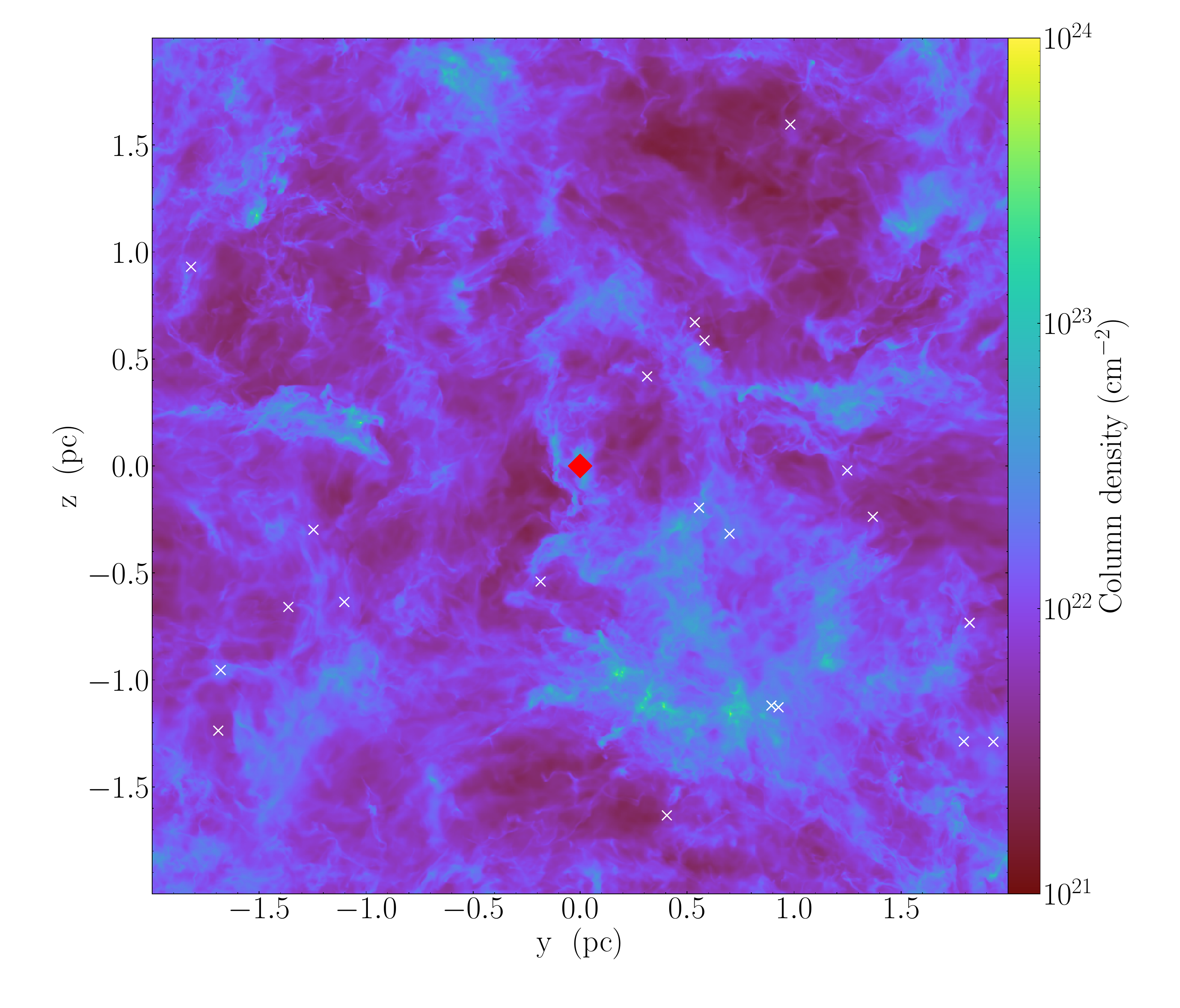}}
  \caption{Projected H$_2$ column densities in the molecular cloud simulation. The white crosses show the location of protostars formed in this snapshot from the simulation (t = 305~kyr). The red diamond marks the protostar that is studied here during the pre-stellar phase. The pre-stellar core studied in this work is centered in the figure for clarity because the box has periodic boundary conditions.}
     \label{fig:ramses}
\end{figure}

The underlying physical simulation is similar to the models presented in \citet{2018ApJ...854...35H}, but the linear resolution is twice higher, and there are eight times more cells overall to resolve the stellar environment and turbulent cascade better. It is a 3D MHD simulation of a 4 pc$^3$ star-forming molecular cloud using the code RAMSES with a number of modifications \citep{2002A&A...385..337T}. Star formation in the cloud is driven by an interplay between supersonic turbulence and gravitational collapse. The supersonic turbulence is maintained by solenoidal driving on the largest scale, resulting in an RMS velocity of 2 km/s in the cloud. The simulation reproduces the empirically derived core mass function and initial mass function to a high degree \citep[]{2018ApJ...854...35H, 2021MNRAS.504.1219P}. The overall characteristic of the cloud simulation is therefore comparable to the observed star formation process. To create realistic initial conditions and erase any trace of the homogeneous initial conditions, the box was first evolved with pure MHD for 20 turbulent turn-over times (20 Myr), and only then were self-gravity and the star formation recipe turned on. 
The simulation uses adaptive mesh refinement, and the smallest cells are 25~au across. The entire model domain is shown in Fig. \ref{fig:ramses}. The RAMSES simulation has already been used in numerous studies to investigate late-stage infall \citep{2023EPJP..138..272K},
chemistry in pre-stellar cores and embedded protostars \citep{2021A&A...649A..66J, 2023A&A...675A..34J}, angular
momentum in Class II systems \citep{2025A&A...694A.327P}, early
Bondi-Hoyle accretion \citep{2025NatAs...9..862P}, and RAMSES-based
zoom-in studies within molecular clouds \citep{2022Natur.606..272J, 2024MNRAS.534.3176T, 2024MNRAS.534.3194A}. 

The model is isothermal and does not consider radiation processes. These are included by post-processing snapshots from the simulation. Each RAMSES snapshot was post-processed using RADMC3D \citep{2012ascl.soft02015D} to estimate the local radiation field within the model. This step was made on a subset of the full simulation domain with a radius of 50,000~au, centered on the pre-stellar core, for computational efficiency. From the RADMC-3D simulations, we derived the dust temperature and the local radiation field, which we used to derive the local visual extinction $A_\mathrm{v,eff}$. We denote this as the effective visual extinction to distinguish it from the line-of-sight extinction as seen by observers. 

The radiation source in the RADMC-3D simulation was the interstellar radiation field (ISRF) surrounding the box. The ISRF was parameterized following \citet{2017A&A...604A..58H}. No internal radiation source was used, as the core is pre-stellar and does not yet harbor a hydrostatic core. The ISRF was attenuated by a visual extinction of the ambient cloud of either $A_\mathrm{v,eff}^\mathrm{amb}$=~1~mag or 2~mag, consistent with simulations presented in \citet{2018A&A...615A..15S} and \citet{2022A&A...668A.131S}. This attenuation was intended to mimic the effects of the core being embedded in a larger ambient molecular cloud that provides additional shielding against the ISRF.

 We assumed perfect thermal coupling between the gas and dust particles. This assumption is valid at higher densities ($\geq$10$^{4-5}$ cm$^{-3}$), which characterize the core itself, but not accurate in the outer diffuse boundary layers \citep{2001ApJ...557..736G}. Nevertheless, the differences between the dust and gas temperatures are expected to remain modest ($\leq$ 2~K) at the edge of cores as well \citep{2017A&A...607A..26S}. 
 
\citet{2023A&A...675A..34J} selected a snapshot that matched the conditions of L1544 best as derived from observations: a central density of $10^6~\mathrm{cm}^{-3}$ within the high-density kernel with a radius of 1800~au \citep{2019ApJ...874...89C}. Because this snapshot occurred 300 kyr after tracer particles were injected and star formation enabled in the simulation, we were able to study the evolution of each tracer particle for a total of 300 kyr before the snapshot in question.
 In Appendix \ref{app:3D_core} we visualize the physical structure of the anisotropic core using eight radial profiles.

\subsubsection{Dynamical evolution tracks using tracer particles}
\begin{figure*}[ht]
\resizebox{\hsize}{!}
        {\includegraphics{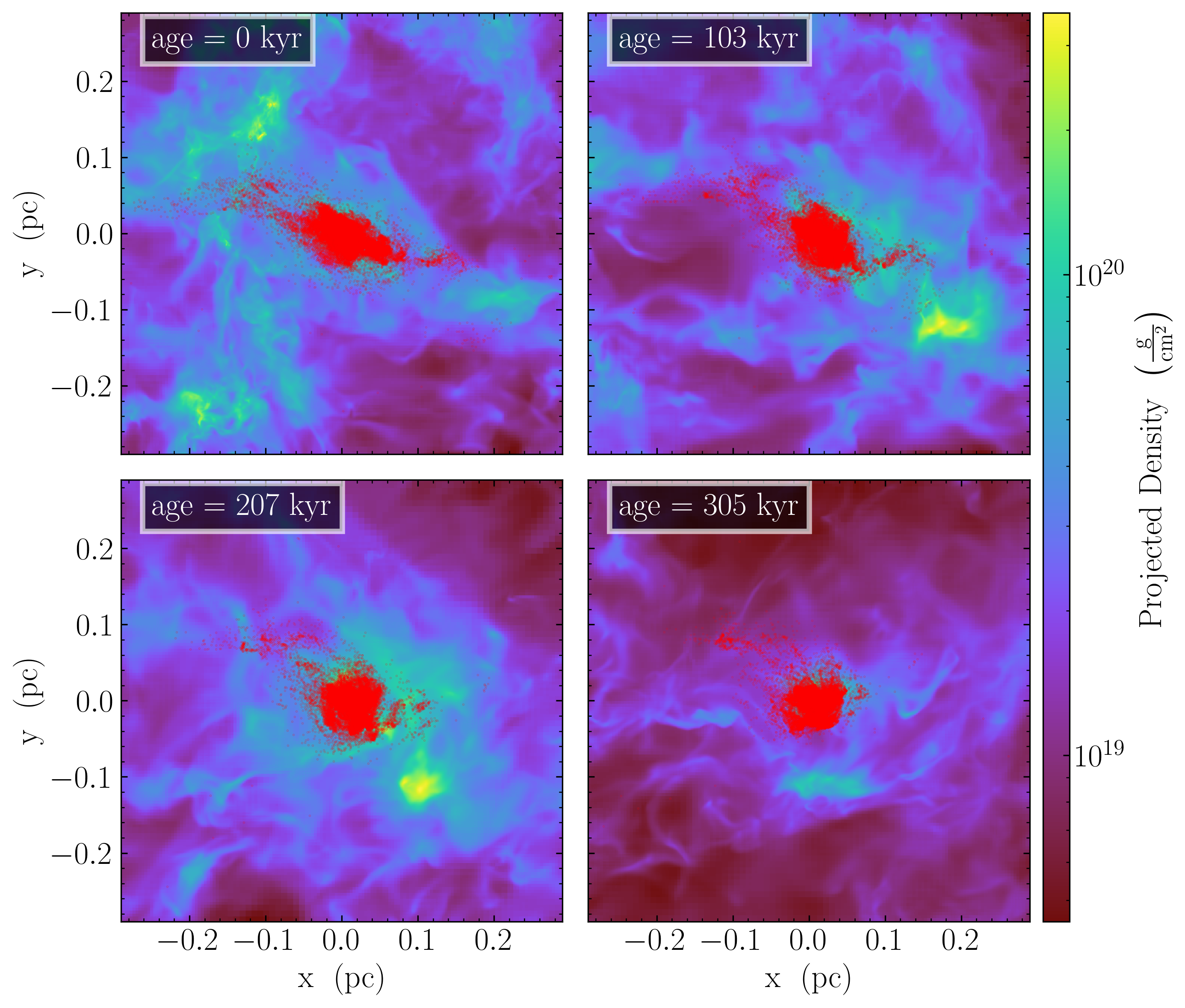}}
  \caption{Four snapshots showing the accretion onto the pre-stellar core at different stages of the simulation. The red dots mark the position of the tracer particle accreted onto the pre-stellar core. Only 20\% of the tracer particles are plotted for clarity. The color map shows the projected column density, similar to Fig. \ref{fig:ramses}. Zero-age is defined as the onset of star formation in the simulation.}
     \label{fig:tracer_4snaps}
\end{figure*}

In the static model of the core, the chemistry was evolved for a given period of time at fixed physical conditions. Such models are broadly used to interpret observations of cores \citep[e.g.,][]{2016ApJ...830L...6J, 2018ApJ...869..165L, 2021A&A...656A.109R, 2025A&A...701A.291R, 2025ApJ...990..163B}. However, given the highly dynamical nature of star-forming regions and the nonspherical accretion processes, it is important to improve our understanding of how the accretion dynamics affect the chemical evolution and the resulting chemical structure.
We used tracer particles to capture the physical evolution of the gas that is accreted onto the core \citep{2020A&A...643A.108C, 2021MNRAS.505.3442F, 2021A&A...649A..66J, 2023MNRAS.523.6138P, 2023MNRAS.524.5971P, 2024A&A...685A.112N}. Tracer particles are passive, massless particles in the MHD simulations that are advected along with the local gas velocity field. During the simulation, the tracer particles record (trace) the local physical conditions, namely the local density and velocity field. The local temperature and visual extinction were derived from the RADMC-3D post-processing as described in the previous section and applied to each tracer particle trajectory. Hence, the trajectories have the necessary information for computing the chemical history: density, temperature, and irradiation. 
Figure \ref{fig:tracer_4snaps} shows the tracer particles (red dots) accreted onto the pre-stellar core in four different snapshots.
Examples of the physical evolution for five tracer particles are shown in Appendix \ref{app:trajectories}.

\subsection{Chemical model}

We employed a two-phase gas-grain chemical model based on the framework presented in \citet{2023A&A...675A..34J} and initially introduced in \citet{2021A&A...649A..66J}.
Most of the chemical network and parameters remained unchanged from previous works, and we here limit ourselves to a brief introduction of the model and the updates applied in this work.
The gas-phase chemical network was based on the network initially presented in \citet{2017MNRAS.466.4470M}, based on KIDA2014 \citep{2015ApJS..217...20W}, with the addition of deuterated species and spin chemistry for light hydrogen-bearing species (H$_2$, H$_{2}^{+}$, H$_{3}^{+}$, and deuterated isotopologs). The gas-phase network was updated to use corrected rates for the H$_3^+$ + H$_2$ reactions following \citet{2017A&A...607A..26S}. Furthermore, photodissociation and photoionization rates were updated based on \citet{2017A&A...602A.105H}. The full network and chemical model is available for the community on GitHub\footnote{https://github.com/ssjensen92/kemimo}.
The grain-surface chemical network was derived from \citet{2015A&A...584A.124F}. We updated the grain-surface formation routes for methanol (CH$_3$OH) and deuterated isotopologs to follow the activation barriers presented in \citet{2023A&A...680A..87R}.
The model considers grain-surface reactions via the Langmuir-Hinshelwood mechanism and desorption through chemical reactions, grain-heating by cosmic rays, and UV photons. Cosmic-ray-induced grain heating is implemented following \citet{1985A&A...144..147L}. Chemical desorption follows the description from \citet{2007A&A...467.1103G} with an efficiency factor of 0.01. For CH$_3$OH, nonthermal desorption primarily proceeds through chemical desorption. The assumed photodesorption yield for CH$_3$OH was $\gamma_\mathrm{PD} = 10^{-6}$, following the results from \citet{2016ApJ...817L..12B} for ice mixtures with CH$_3$OH and CO.
 The gas-phase network included the same reaction types as in the standard KIDA network (three-body reactions were excluded). The model included self-shielding of H$_2$, CO, HD, and N$_2$ \citep{2009A&A...503..323V, 2011MNRAS.412.2603W, 2011MNRAS.418..838W, 2014A&A...562A..61H}. The column densities used to calculate the self-shielding factors were not computed self-consistently due to computational constraints. Instead, $N(\mathrm{H}$) was derived from the observational relation $N(H) = A_\mathrm{v}  / 1.59\times10^{21}~\mathrm{cm}^{-2}$ \citep{1994ApJ...427..274D}, where we assumed $A_\mathrm{v} \approx A_\mathrm{v,eff}$. To estimate the N$_2$ and CO self-shielding factors, the local gas-phase ratio with respect to H$_2$ was used to estimate the column densities.
The complete network contained $\sim$950 species and $\sim$50,000 reactions.

To improve the self-consistency of the model, we adopted the parametric description of the cosmic-ray attenuation as a function of H$_2$ column density presented in \citet{2018A&A...614A.111P}. Specifically, we opted for the low model. The H$_2$ column density was computed from the local visual extinction, $A_\mathrm{v,eff}$, similarly to the approach described for the self-shielding factors. In the previous model, presented in \citet{2023A&A...675A..34J}, a fixed cosmic-ray ionization rate of $\zeta = 1.3\times10^{-17}$~s$^{-1}$ was assumed throughout the core.

We ran models with two different sets of initial elemental abundances, as listed in Table \ref{tab:abun}. One set of abundances was similar to the abundances used in \citet{2023A&A...675A..34J} and is labeled LM for ``low metal.'' The second set contained the high-metal (HM) abundances from \citet[labeled EA2, ][]{2008ApJ...680..371W}. These initial abundances are closer to the cosmic abundances and represent the abundances expected at the earliest stages of molecular clouds, prior to any chemical evolution, which may cause depletion of heavier elements (S, Si, Mg, Fe, and Cl) in the gas phase due to accretion onto dust grains or incorporation into larger molecules.

For the chemical simulations, several different initial conditions were tested. Our fiducial model included a molecular cloud phase with static physical conditions for 1~Myr with $n_\mathrm{H}$=~1000~cm$^{-3}$, $T_\mathrm{gas}$= $T_\mathrm{dust}$=~14~K, and $A_\mathrm{v,eff}^\mathrm{amb}$=~2~mag. Other models were tested with a lower ambient extinction of 1~mag, higher $T_\mathrm{gas/dust} = 15$~K, and different durations of this static phase, including models without any static phase before the tracer particle evolution. After the initial static phase, the tracer particle phase was computed for 300 kyr of the tracer trajectories. The ambient extinction assumed during the initial phase was also applied during the tracer particle phase, that is, $A_\mathrm{v,eff}$=$A_\mathrm{v,eff}^\mathrm{amb}$+$A_\mathrm{v,eff}$.
One set of models was also run with a higher density of $n_\mathrm{H}$=~10$^{4}$~cm$^{-3}$, a higher ambient extinction of $A_\mathrm{v,eff}^\mathrm{amb}$=~4~mag, and a lower temperature of 10~K. For this model, the ambient extinction for the tracer particle trajectories remained at $A_\mathrm{v,eff}^\mathrm{amb}$ = 2~mag, which provides a better fit with the predicted temperature profile of L1544 (see Fig. \ref{fig:temp_profiles}). An overview of the different initial conditions is presented in Table \ref{tab:chemmodels}.

\begin{table}
\caption{Initial elemental abundances for the LM and HM models.}
\label{tab:abun}
\centering
\begin{tabular}{lcc}
\hline
Species & LM & HM \\ \hline
o-H$_2$ &      4.995($-$1)          &  4.995($-$1) \\
p-H$_2$ &      5.00($-$4)           & 5.00($-$4)  \\
HD      &      1.60($-$5)           & 1.60($-$5)   \\      
He      &      9.00($-$2)           & 9.00($-$2) \\
O       &      2.56($-$4)           &  2.56($-$4)\\
C$^{+}$ &      1.20($-$4)           & 1.20($-$4)   \\
N       &      7.60($-$5)           &  7.60($-$5)      \\
S$^{+}$ &      8.00($-$8)           &  1.50($-$5)     \\ 
Si$^{+}$   & 8.00($-$9)         & 1.70($-$6)          \\ 
Mg$^{+}$   & 7.00($-$9)         & 2.40($-$6)   \\ 
Fe$^{+}$   & 3.00($-$9)         & 2.00($-$7)       \\ 
Na$^{+}$   & 2.25($-$9)         & 2.00($-$7)          \\ 
Cl$^{+}$   & 1.00($-$9)         & 1.8($-$8)          \\ 
 F$^{+}$   & 6.68($-$9)        & 1.8($-$8)\\ 
  P$^{+}$   & 2.00($-$10)        & 1.2($-$7)\\ \hline
A(B) = A~$\times$~10$^{\mathrm{B}}$.
\end{tabular}
\tablefoot{ An initial H$_2$ ortho-to-para ratio of 10$^{-3}$ was assumed.}
\end{table}

\begin{table}
\caption{Initial conditions for the chemical simulation before the calculation of the evolution during the tracer particle phase.}
\label{tab:chemmodels}
\centering
\begin{tabular}{lrccc}
\hline
Model & $n_\mathrm{H}$(cm$^{-3}$) & $A_\mathrm{v,eff}^\mathrm{amb}$(mag) & $T_\mathrm{gas}$(K) & $t_\mathrm{init}$(Myr) \\ \hline
\#1 & 10$^{3}$ & 2.0  & 14 & 0.0 \\ 
\#2 & 10$^{3}$ & 2.0  & 14 & 0.5 \\ 
\#3 & 10$^{3}$ & 2.0  & 14 & 1.0 \\ 
\#4 & 10$^{3}$ & 2.0  & 14 & 2.0 \\ 
\#5 & 5$\times$10$^{2}$ & 1.0  & 15 & 0.0 \\ 
\#6 & 5$\times$10$^{2}$ & 1.0  & 15 & 0.5 \\ 
\#7 & 5$\times$10$^{2}$ & 1.0  & 15 & 1.0 \\ 
\#8 & 5$\times$10$^{2}$ & 1.0  & 15 & 2.0 \\ 
\#9 & 10$^{4}$ & 4.0 & 10 & 1.0 \\ \hline
\end{tabular}
\tablefoot{All models were run with LM and HM initial abundances.}
\end{table}


\subsection{Radiative transfer modeling: Synthetic images and spectral cubes}
The final step of the model presented here involved the nonlocal thermodynamic equilibrium  radiative transfer simulation using the code {\sc LIME} \citep{2010A&A...523A..25B}. The computation of the radiative transfer for selected molecular transitions allowed us to directly compare the model with observations of L1544. The {\sc LIME} models solve the radiative transfer equation in a domain with a radius of $\sim$50,000 au around the core. The physical structure of the core was based on the {\sc RAMSES} and {\sc RADMC-3D} simulations, which provided the density, velocity, and temperature fields. Thermal turbulence was included self-consistently for each molecule based on the local gas temperature.
As in the chemical modeling, we assumed $T_\mathrm{gas} = T_\mathrm{dust}$. 
The collisional rates for the molecules presented here were accessed through the LAMDA \citep{2005A&A...432..369S}\footnote{https://home.strw.leidenuniv.nl/~moldata/} and EMEAA\footnote{https://emaa.osug.fr/} databases. A complete list of the references for the collisional rates is provided in Appendix \ref{app:1}.
The output of {\sc LIME} models are spectral cubes with a predefined bandwidth and spectral resolution. The distance to the modeled core was set to 170~pc, similar to the distance to L1544 \citep{2019A&A...630A.137G}. The spectral cube was convolved with a Gaussian beam with a beam width similar to that of the IRAM 30m telescope ($\sim$ 30$''$) that was used to observe L1544 in \citet{2016A&A...592L..11S} and \citet{2025ApJ...992..124F}.
Hence, a direct comparison between the convolved spectral cube and observations of L1544 is possible.

\section{Results} \label{sec:results}

\subsection{Studying the chemical morphology of $c$-C$_3$H$_2$ and CH$_3$OH}
Multiple studies have revealed distinct morphological patterns among different molecular groups in young cores \citep{2017A&A...606A..82S, 2020A&A...643A..60S}. These spatial differences appear to be linked to the formation or destruction mechanism of the molecules. A key example is the contrast between methanol and carbon-chain molecules, which follow fundamentally different formation pathways: methanol forms on dust grain surfaces via hydrogenation of CO ice, while carbon-chain molecules are primarily synthesized in the gas phase from atomic carbon and are typically suppressed when the majority of carbon is converted into CO.
Following \citet{2017A&A...606A..82S}, we relied on $c$-C$_3$H$_2$ and CH$_3$OH as tracers of carbon-chain and COMs, respectively. 

\begin{figure*}[ht]
\resizebox{\hsize}{!}
        {\includegraphics{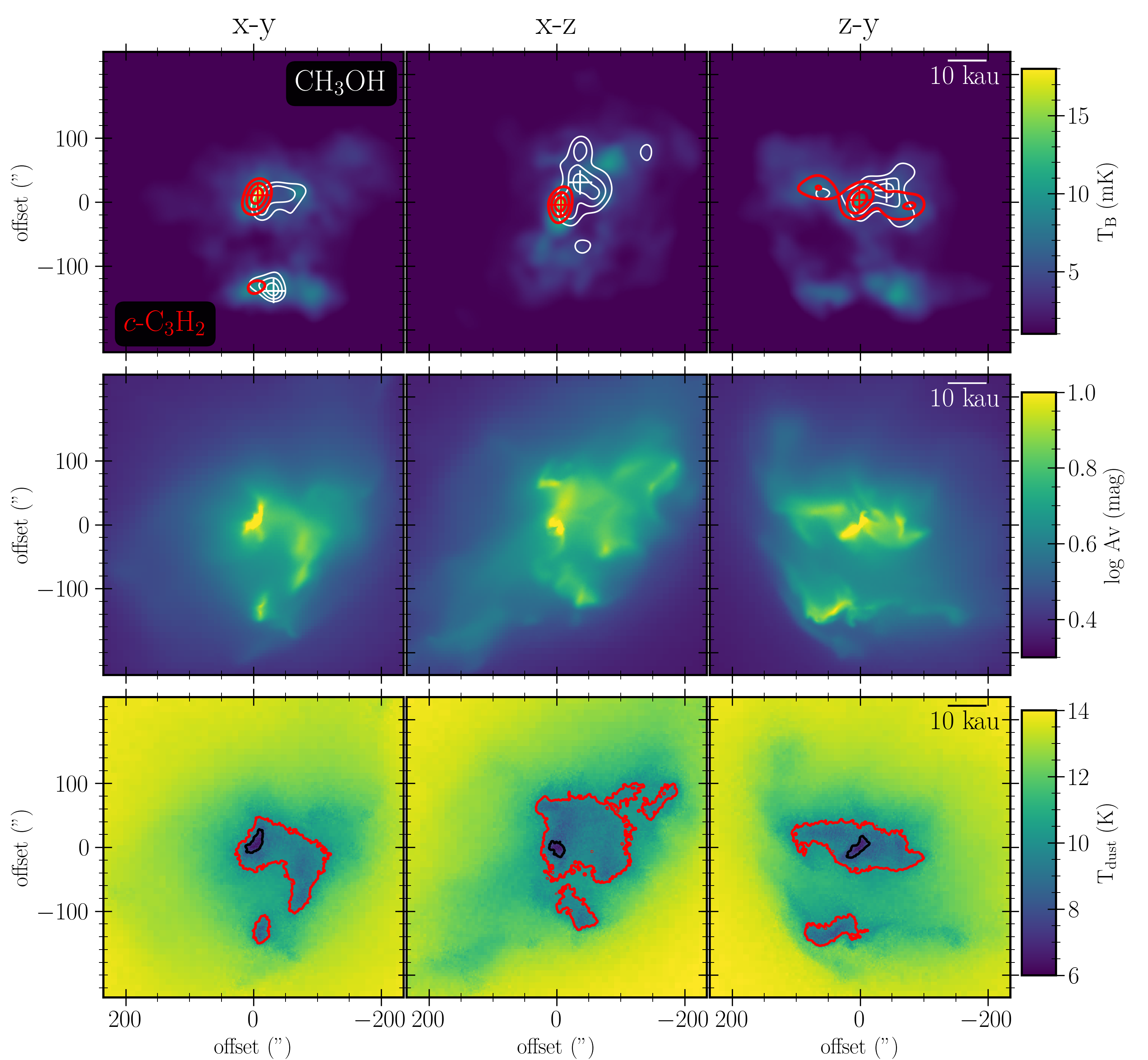}}
  \caption{LM model with 10$^{6}$~yr static phase and $A_\mathrm{v,eff}^\mathrm{amb} = 2$~mag (\#3 LM).
  \emph{Top:} Integrated intensity levels for CH$_3$OH (white contours) and $c$-C$_3$H$_2$ (red contours) along the three principal axes of the simulation. The color map shows the continuum emission at 1.1~mm. 
  \emph{Middle:} Slice plot showing the local $\log$ $A_\mathrm{v,eff}$ for each principal axes. The upper range of the color map is limited to 10~mag for clarity.
                \emph{Bottom:} Slice plot showing the local dust temperature for each principal axes. The red and black contours show the 10~K and 7~K limits, respectively.
  }
     \label{fig:chem2plot_fidHM}
\end{figure*}

\begin{figure*}[ht]
\resizebox{\hsize}{!}
        {\includegraphics{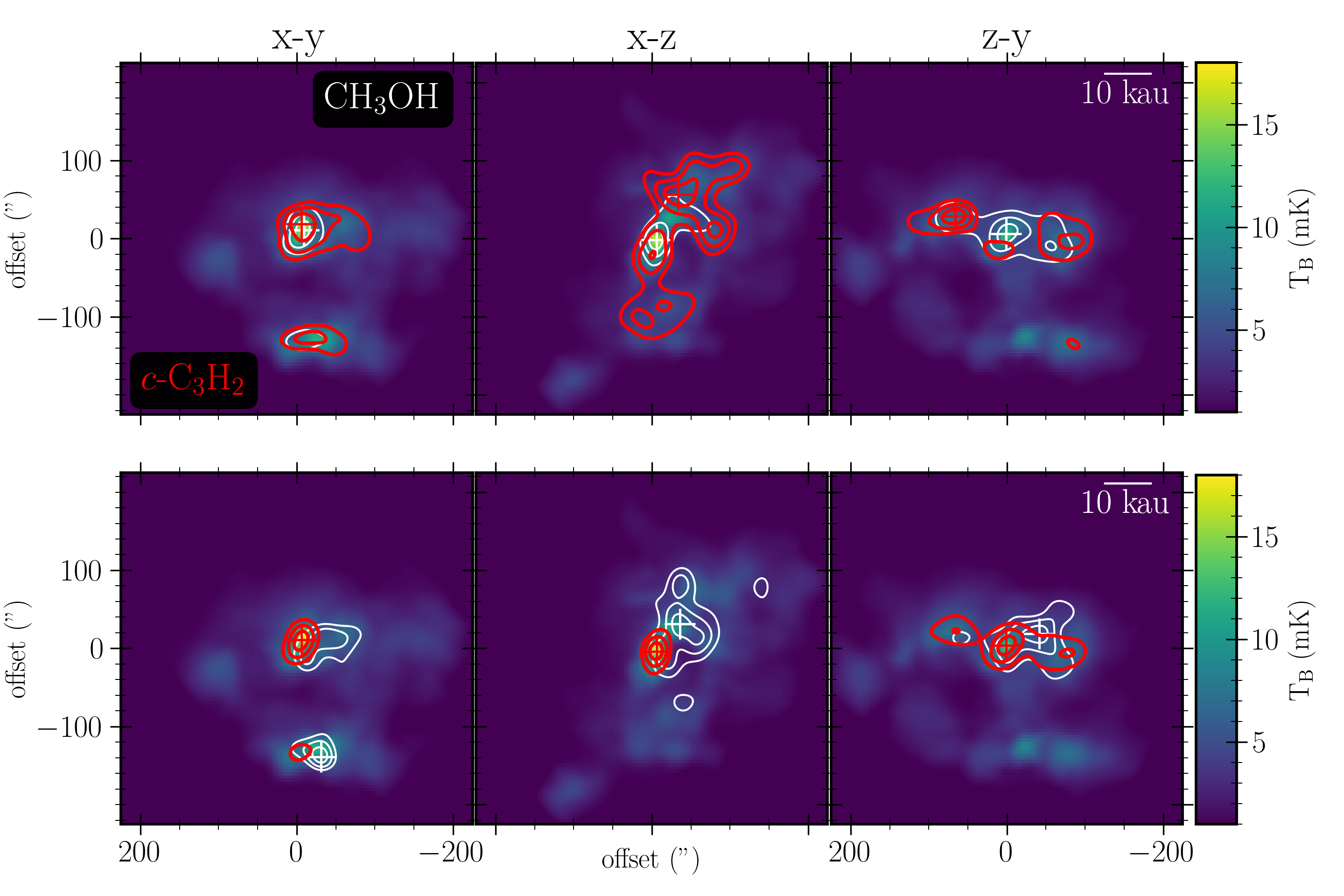}}
  \caption{ Comparison of the fiducial static model and the fiducial dynamic model, both with the updated network from this work.
  \emph{Top:} Static model with LM abundances. The contours show the integrated-intensity levels for CH$_3$OH (white) and $c$-C$_3$H$_2$ (red) along the three principal axes of the simulation. The color map shows the continuum emission at 1.1~mm. 
                \emph{Bottom:} Similar to the top panel but for dynamical model $\#7$ LM.
  }
     \label{fig:chem2plot_staticdyn}
\end{figure*}

Figure \ref{fig:chem2plot_fidHM} shows a comparison between the molecular emission of $c$-C$_3$H$_2$ and CH$_3$OH and the physical structure of the core for model \#3 with higher metallicities. The figure divides the model into three principal planes ($x$-$y$, $x$-$z$, and $z$-$y$). The transitions modeled in this figure are identical to those observed in \citet{2016A&A...592L..11S, 2017A&A...606A..82S}. We can therefore directly compare the observed morphologies and intensities.
In the $x$-$y$ plane and $x$-$z$ planes, the peak emission and contours of CH$_3$OH are offset from the dust-continuum peak position, while the $c$-C$_3$H$_2$ emission peak coincides with the dust peak. The offsets of CH$_3$OH coincide with the more shielded regions of the core, as shown by the $A_\mathrm{v,eff}$ and $T_\mathrm{dust}$ slice plots. This is consistent with the morphology observed toward L1544 for these two species in L1544. In the third principal plane, no clear offset is seen, and the contours of both molecules trace the denser and more shielded region of the core.
The static model presented in \citet{2023A&A...675A..34J} was able to qualitatively reproduce the observed morphology in L1544 for $c$-C$_3$H$_2$ and CH$_3$OH. Given the updated chemical model in this work, we reran the static model and compared the new results with the dynamical model presented in this work. Figure \ref{fig:chem2plot_staticdyn} shows the morphology of the static model with the updated chemical model in the top panel. The static model has changed morphology since the one presented in \citet{2023A&A...675A..34J}: the methanol emission is now centered in the dust continuum peak in all planes. This change is primarily due to the updated methanol formation pathways, which include new activation energies from \citet{2023A&A...680A..87R}, one additional abstraction reaction, and the correction of branching ratios, which have altered the efficiency of methanol formation in the high-density kernel of the core.
The bottom panel shows the results for dynamical model \#3 with standard metallicities. Similarly to the model presented in Fig. \ref{fig:chem2plot_fidHM}, the CH$_3$OH emission is offset from the dust peak toward the regions with higher extinction. 

A direct comparison between static and dynamical models under identical conditions is not directly possible: the static model was evolved for a fixed time at fixed physical conditions, while the dynamical model considered an initial static phase followed by $\sim$300~kyr of dynamical evolution as the gas accretes toward the core. In Fig. \ref{fig:chem2plot_staticdyn} we chose to compare the fiducial model presented in \citet[$t_\mathrm{static} = 10^6$~yr, $A_\mathrm{v,eff}^\mathrm{amb}$=~2~mag]{2023A&A...675A..34J} with model $\#3$~LM in this work ($t_\mathrm{init}=10^{6}$~yr, $A_\mathrm{v,eff}^\mathrm{amb}$=~2~mag), but we note that the two models are fundamentally different in terms of physical and chemical evolution. In the static model, $t_\mathrm{static}$ denotes the duration of the chemical simulation. Hence, in the fiducial model from \citet{2023A&A...675A..34J}, the chemical evolution was modeled for a duration of 1~Myr.

\begin{figure*}[ht]
\resizebox{\hsize}{!}
        {\includegraphics{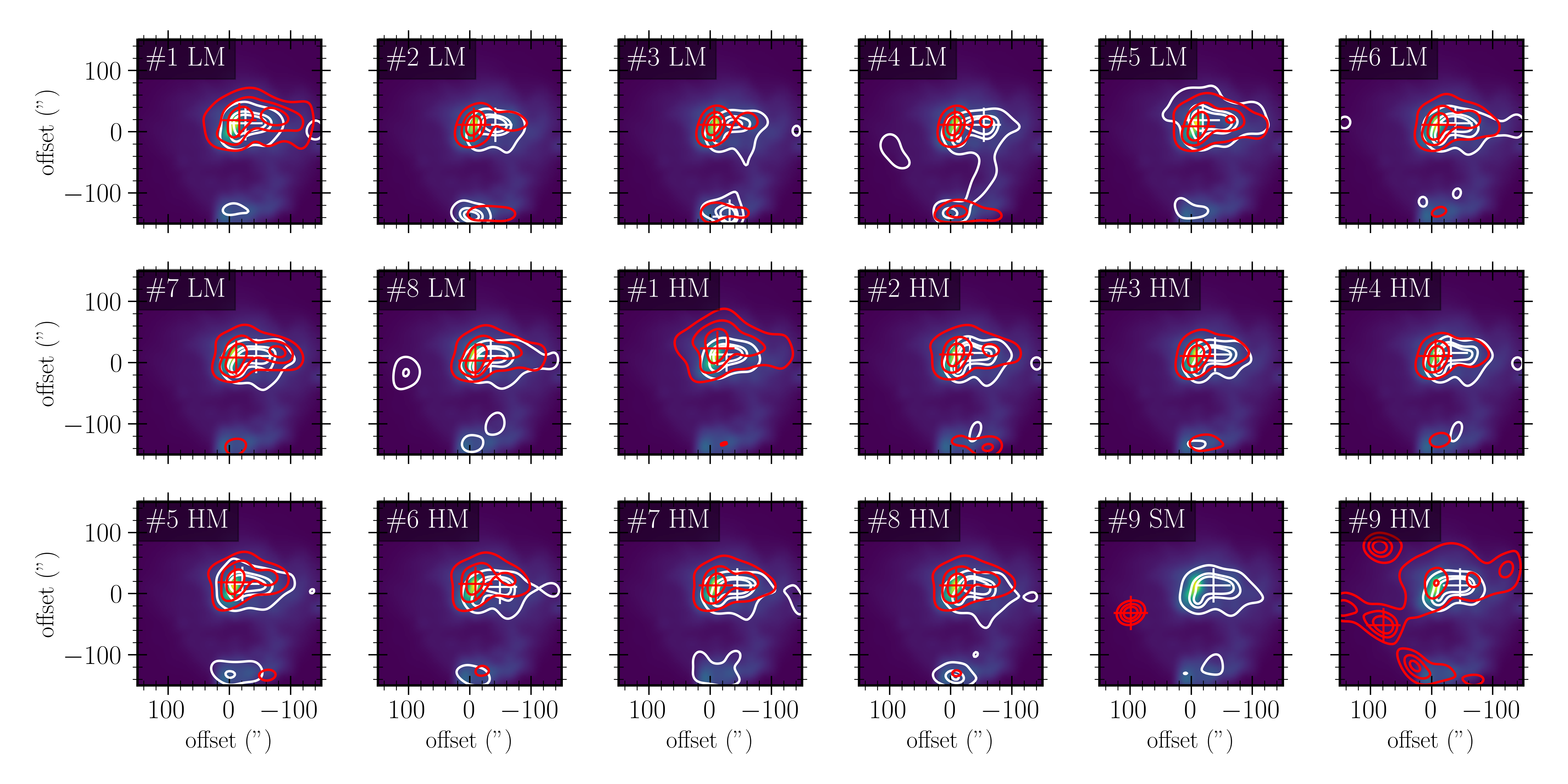}}
  \caption{ Comparison of the 18 dynamical models. The red and white contours indicate the integrated emission contours for $c$-C$_3$H$_2$ and CH$_3$OH, respectively. The background shows the continuum map of the core at 1.1~mm. The $c$-C$_3$H$_2$ emission generally peaks at the dust continuum peak, while CH$_3$OH is offset toward the more shielded part of the core.
  }
     \label{fig:allmodels}
\end{figure*}

An overview of the morphologies in all 18 dynamical models is presented in Fig. \ref{fig:allmodels}. Only two models show a significant difference for the morphologies: model \#9 LM and HM show a different morphology for $c$-C$_3$H$_2$, which under these conditions traces less dense clumps of gas that are offset from the core itself.

\begin{figure}[ht]
\resizebox{\hsize}{!}
        {\includegraphics{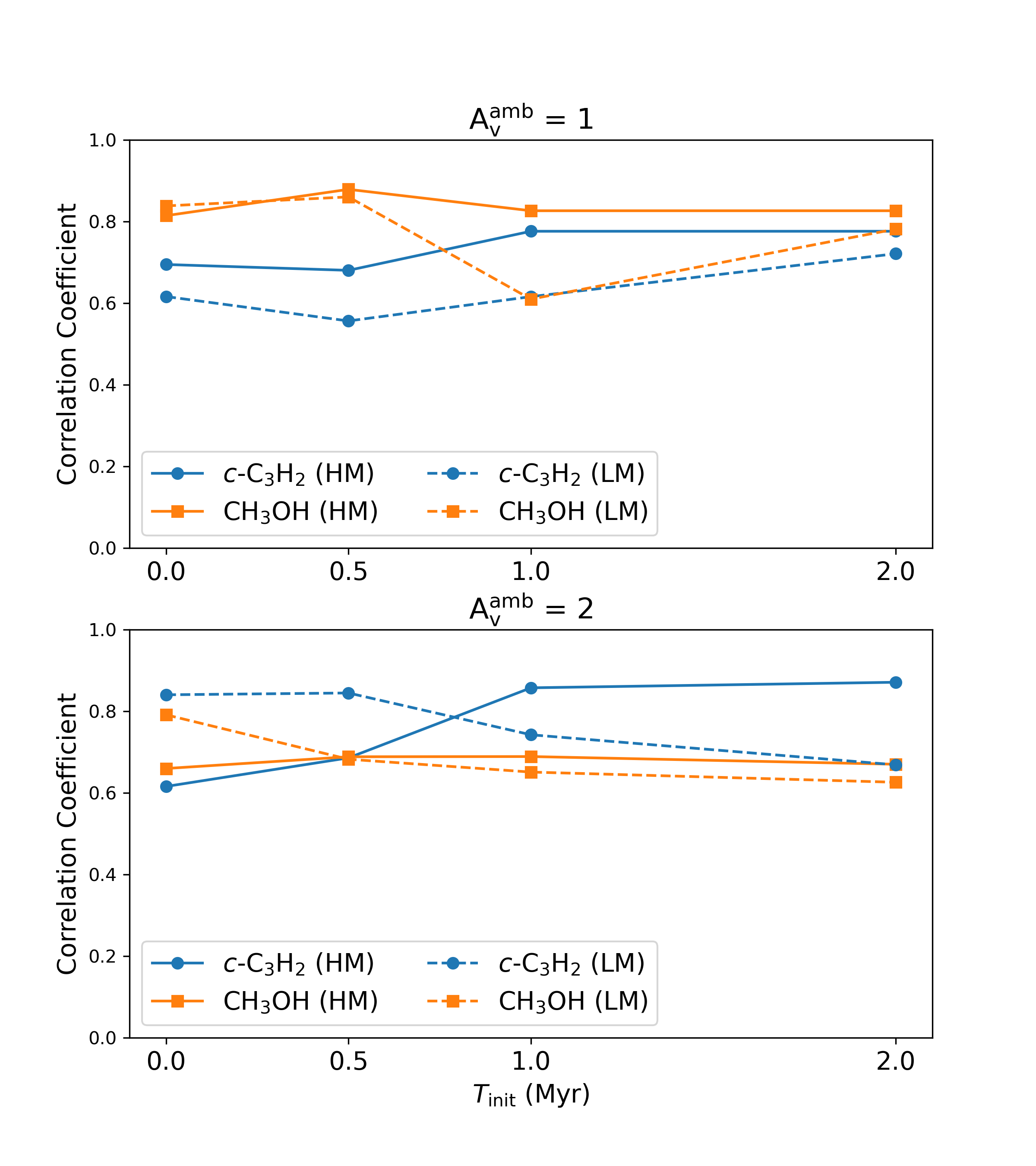}}
  \caption{ Correlation between the column densities of $c$-C$_3$H$_2$, CH$_3$OH, and $A_\mathrm{v,eff}$ for models \#1--\#8 for standard and higher metallicities. With $A_\mathrm{v,eff}^\mathrm{amb}$=~1~mag,  CH$_3$OH shows a stronger correlation, especially at lower $t_\mathrm{init}$. For $A_\mathrm{v,eff}^\mathrm{amb}$=~2~mag, the correlation is lower for CH$_3$OH and stronger for $c$-C$_3$H$_2$.
  }
     \label{fig:corr_time}
\end{figure}

To quantify the observed patterns, we computed the column density maps for the two molecules in each plane and compared them with the integrated column of H$_2$, $A_\mathrm{v,eff}$, and $T_\mathrm{gas}$ \footnote{The integrated columns for the visual extinction $A_\mathrm{v,eff}$ and $T_\mathrm{gas}$ do not have a direct physical meaning here, but represent the structure of the core along the line of sight.}. While the synthetic maps from the {\sc LIME} models allow a direct comparison between observations, studying the column densities provides an alternative view of the underlying chemical structure along the line of sight, independent of excitation effects. 
To determine to which degree the different molecules correlate with the physical conditions along the line of sight, we computed the Pearson correlation coefficients for the molecular column densities and the visual extinction, local gas density, and the gas temperature. The results for two models are reported in Tables \ref{tab:pearson_n1e3_T1e6_Av2_SM_P18} and \ref{tab:pearson_n5e2_T1e6_Av1_SM_P18}. As expected, the results show a strong correlation between $N_{\mathrm{H}_2}$ and the integrated column of the visual extinction $\hat{N}(A_\mathrm{v,eff})$, while an anticorrelation is seen for the integrated column of the gas temperature. 
In general, $N$(CH$_3$OH) and $N$($c$-C$_3$H$_2$) show positive correlations with $N$(H$_2$) and $\hat{N}(A_\mathrm{v,eff})$. For CH$_3$OH, this is expected because the formation pathway depends on the freeze-out of CO, which requires higher densities and higher extinction. For $c$-C$_3$H$_2$, the results are more surprising because the formation of $c$-C$_3$H$_2$ also occurs efficiently at lower densities and extinction.
The correlation coefficients between $N$(CH$_3$OH) and $N$($c$-C$_3$H$_2$) with $\hat{N}(A_\mathrm{v,eff})$ are shown in Fig. \ref{fig:corr_time} as a function of $t_\mathrm{init}$. A general trend is present in the correlation coefficients between $c$-C$_3$H$_2$, CH$_3$OH, and $A_\mathrm{v,eff}$. The models with lower $A_\mathrm{v,eff}^\mathrm{amb}$=~1~mag (\#5-\#8) show a stronger correlation between methanol and the integrated extinction ($\hat{N}(A_\mathrm{v,eff})$), while the correlation is weaker for models with $A_\mathrm{v,eff}^\mathrm{amb}$=~2~mag (\#1-\#4). Additionally, the correlation is more pronounced in models with a shorter initial phase, representing a younger chemistry. The trend is visualized in Fig. \ref{fig:corr_time}. We interpret this trend as a result of the minimum requirements for CH$_3$OH formation. Tracer particles need sufficient time at higher density and $A_\mathrm{v,eff}$ ($n$(H) $\gtrsim10^{4}$, $A_\mathrm{v,eff} \gtrsim$~3) for CH$_3$OH formation to proceed. In models with lower $A_\mathrm{v,eff}^\mathrm{amb}$, there is a stronger dependence on the individual tracer particle trajectories. Similarly, in models with a shorter $t_\mathrm{static}$, the dependence on the individual tracer particle trajectories is enhanced. 

It should be noted that the abundance of $c$-C$_3$H$_2$ is very low in models with a more evolved chemistry (\#2-\#4). Hence, the correlation with high $\hat{N}(A_\mathrm{v,eff})$ and $N_{\mathrm{H}_2}$ is a result of the very weak emission that is confined to the region with the highest densities. For methanol, on the other hand, this is not the case. Methanol is weaker in the models with a younger chemistry (\#5-\#6), which shows the strongest correlation with the integrated H$_2$ column density and $\hat{N}(A_\mathrm{v,eff})$.

\begin{table*}
\caption{Pearson correlation coefficients for model \#3 LM in ($x$-$y$, $x$-$z$, $z$-$y$) planes.}
\centering
\begin{tabular}{l c c c c c }
\hline\hline
  & $N$(H$_2$) & $\hat{N}(A_\mathrm{v,eff})$ & $\hat{N}(T_\mathrm{gas})$
   & $N$(CH$_3$OH) & $N$($c$-C$_3$H$_2$) \\
  \hline
$N$(H$_2$) & -- & (0.75, 0.77, 0.70) & (-0.70, -0.73, -0.65) & (0.53, 0.49, 0.49) & (0.41, 0.43, 0.40) \\
$\hat{N}(A_\mathrm{v,eff})$ & (0.75, 0.77, 0.70) & -- & (-0.98, -0.99, -0.99) & (0.61, 0.67, 0.67) & (0.78, 0.72, 0.73) \\
$\hat{N}(T_\mathrm{gas})$ & (-0.70, -0.73, -0.65) & (-0.98, -0.99, -0.99) & -- & (-0.62, -0.69, -0.68) & (-0.81, -0.75, -0.75) \\
$N$(CH$_3$OH) & (0.53, 0.49, 0.49) & (0.61, 0.67, 0.67) & (-0.62, -0.69, -0.68) & -- & (0.67, 0.71, 0.63) \\
$N$($c$-C$_3$H$_2$) & (0.41, 0.43, 0.40) & (0.78, 0.72, 0.73) & (-0.81, -0.75, -0.75) & (0.67, 0.71, 0.63) & -- \\
\hline
\end{tabular}

\label{tab:pearson_n1e3_T1e6_Av2_SM_P18}
\end{table*}

\begin{table*}
\caption{Pearson correlation coefficients for model \#6 LM in ($x$-$y$, $x$-$z$, $z$-$y$) planes.}
\centering
\begin{tabular}{l c c c c c }
\hline\hline
  & $N$(H$_2$) & $\hat{N}(A_\mathrm{v,eff})$ & $\hat{N}(T_\mathrm{gas})$
   & $N$(CH$_3$OH) & $N$($c$-C$_3$H$_2$) \\
  \hline
$N$(H$_2$) & -- & (0.75, 0.77, 0.70) & (-0.70, -0.73, -0.65) & (0.47, 0.41, 0.53) & (0.31, 0.39, 0.39) \\
$\hat{N}(A_\mathrm{v,eff})$ & (0.75, 0.77, 0.70) & -- & (-0.98, -0.99, -0.99) & (0.60, 0.49, 0.74) & (0.55, 0.63, 0.66) \\
$\hat{N}(T_\mathrm{gas})$ & (-0.70, -0.73, -0.65) & (-0.98, -0.99, -0.99) & -- & (-0.60, -0.47, -0.73) & (-0.52, -0.60, -0.62) \\
$N$(CH$_3$OH) & (0.47, 0.41, 0.53) & (0.60, 0.49, 0.74) & (-0.60, -0.47, -0.73) & -- & (0.35, 0.50, 0.48) \\
$N$($c$-C$_3$H$_2$) & (0.31, 0.39, 0.39) & (0.55, 0.63, 0.66) & (-0.52, -0.60, -0.62) & (0.35, 0.50, 0.48) & -- \\
\hline
\end{tabular}

\label{tab:pearson_n5e2_T1e6_Av1_SM_P18}
\end{table*}

\subsection{Comparing the molecular intensities of L1544 and the model}

\begin{figure}[ht]
\resizebox{\hsize}{!}
        {\includegraphics{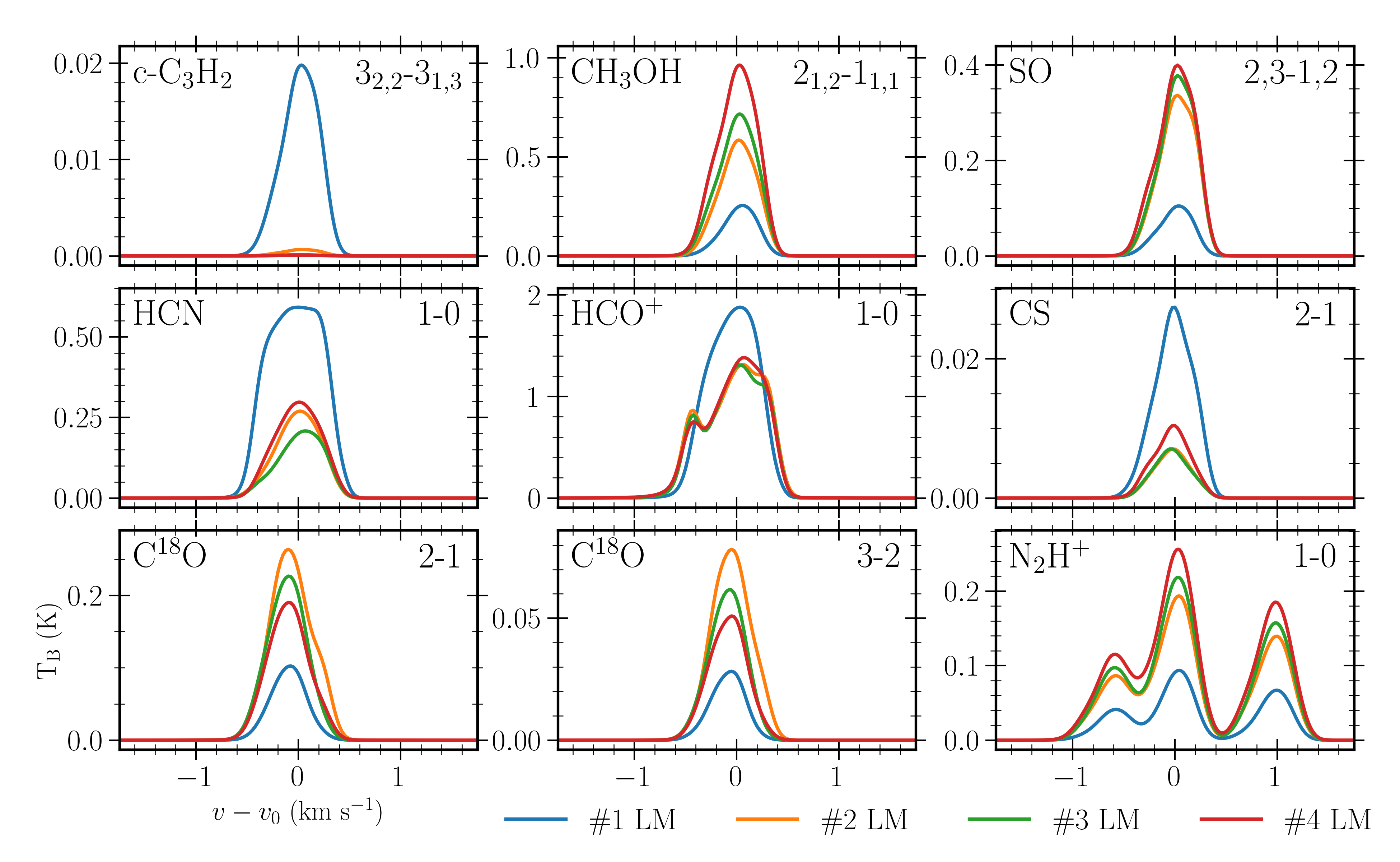}}
  \caption{Line profiles as predicted by the fiducial model with LM abundances, 10$^{6}$~yr static phase, and $A_\mathrm{v,eff}^\mathrm{amb} = 2$~mag.
The spectra were extracted toward the continuum peak in the $x$-$y$ plane. A fixed $^{16}$O/$^{18}$O ratio of 560 was used for the C$^{18}$O abundance profiles.}
     \label{fig:fid_spectra_SM_xy}
\end{figure}

\begin{figure}[ht]
\resizebox{\hsize}{!}
        {\includegraphics{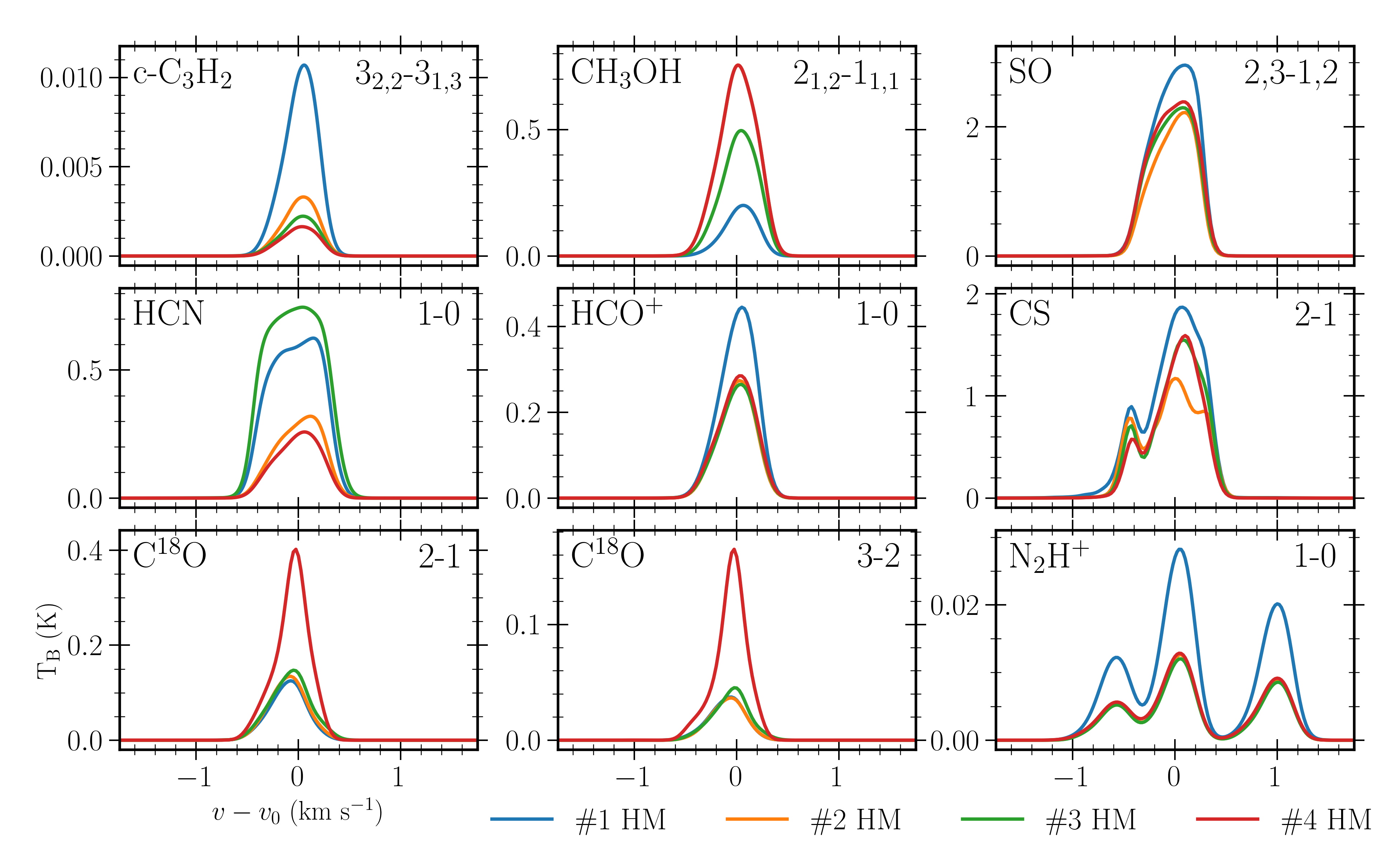}}
  \caption{Line profiles as predicted by the fiducial model with HM abundances, 10$^{6}$~yr static phase, and $A_\mathrm{v,eff}^\mathrm{amb} = 2$~mag.
The spectra were extracted toward the continuum peak in the $x$-$y$ plane. A fixed $^{16}$O/$^{18}$O ratio of 560 was used for the C$^{18}$O abundance profiles.}
     \label{fig:fid_spectra_HM_xy}
\end{figure}

Figures \ref{fig:fid_spectra_SM_xy} and \ref{fig:fid_spectra_HM_xy} shows synthetic spectra for selected molecules: $c$-C$_3$H$_2$, CH$_3$OH, SO, HCN, HCO$^{+}$, CS,  C$^{18}$O, and N$_2$H$^+$. All spectra were extracted toward the peak in the dust continuum in the $x$-$y$ plane. The behavior of the lines was qualitatively the same in all planes.

To evaluate how well the model reproduces the observed molecular intensities toward the dust peak of L1544, we focused not only on absolute values, but primarily on the relative differences between species. Achieving exact agreement in absolute intensities, even with a chemically perfect model, would require the physical model to precisely replicate the density structure, temperature, irradiation environment, and evolutionary stage of L1544. Because the modeled core differs from the real L1544 in these aspects, a one-to-one comparison of absolute intensities is not meaningful scientifically. Instead, by examining the relative differences in intensities for the different molecules, we identified the species that are simultaneously well reproduced by the model and those that require further refinement, offering a more robust and insightful assessment of the model performance.
For L1544, we compared our results with the observations of CS, SO, and HCO$^{+}$ reported in \citet{2025ApJ...992..124F}, $c$-C$_3$H$_2$ and CH$_3$OH reported in \citet{2016A&A...592L..11S}, N$_2$H$^+$ reported in \citet{2019A&A...629A..15R}, and HCN (priv. comm.). All observations were carried out with the IRAM 30m telescope at $\sim$3~mm.

\begin{figure}[ht]
\resizebox{\hsize}{!}
        {\includegraphics{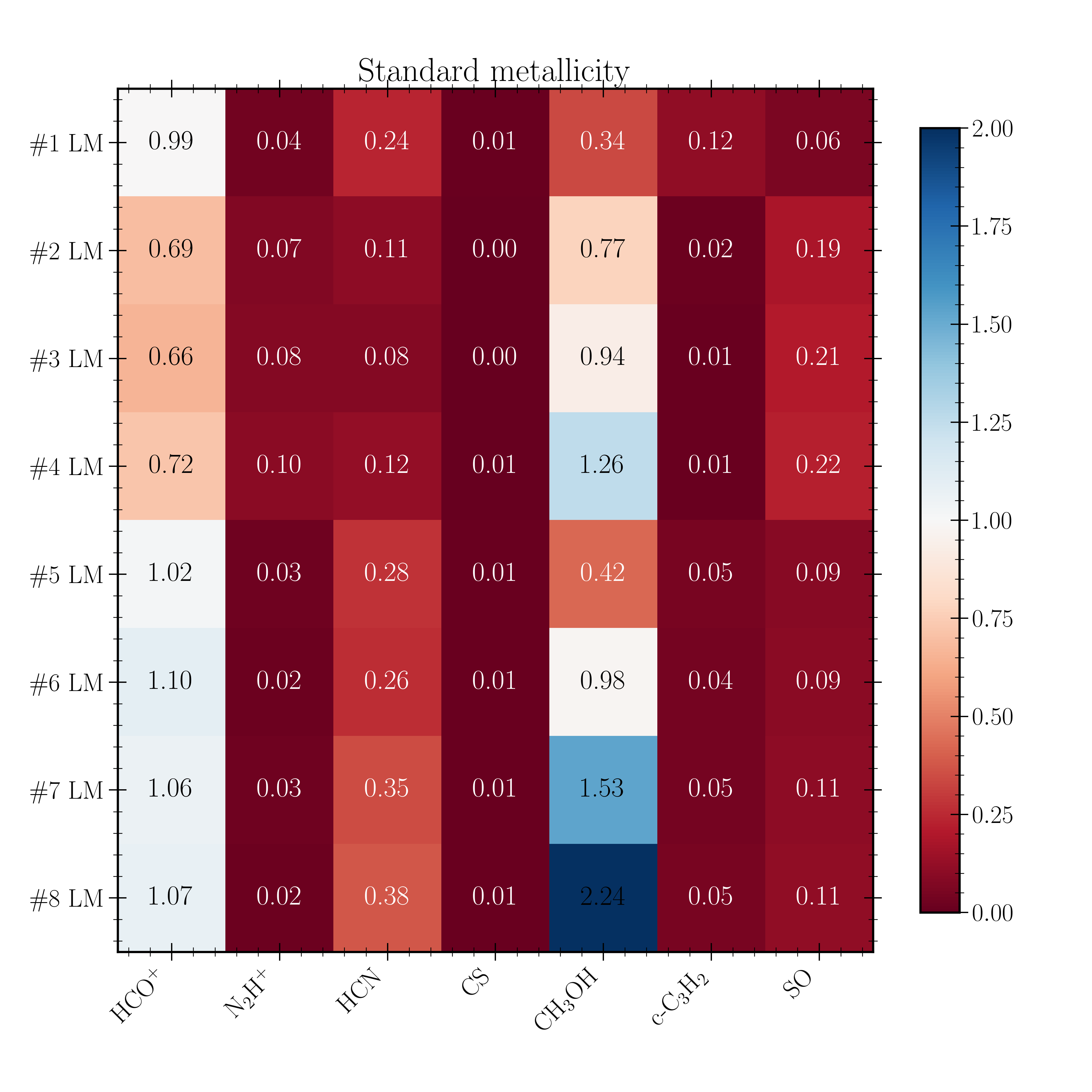}}
  \caption{Comparison table for the peak intensity of the modeled emission lines toward L1544 (dust peak position) and toward the dust peak in the $x$-$y$ plane. This figure shows the comparison for models with standard metallicities.}
     \label{fig:ratio_xy_SM}
\end{figure}

\begin{figure}[ht]
\resizebox{\hsize}{!}
        {\includegraphics{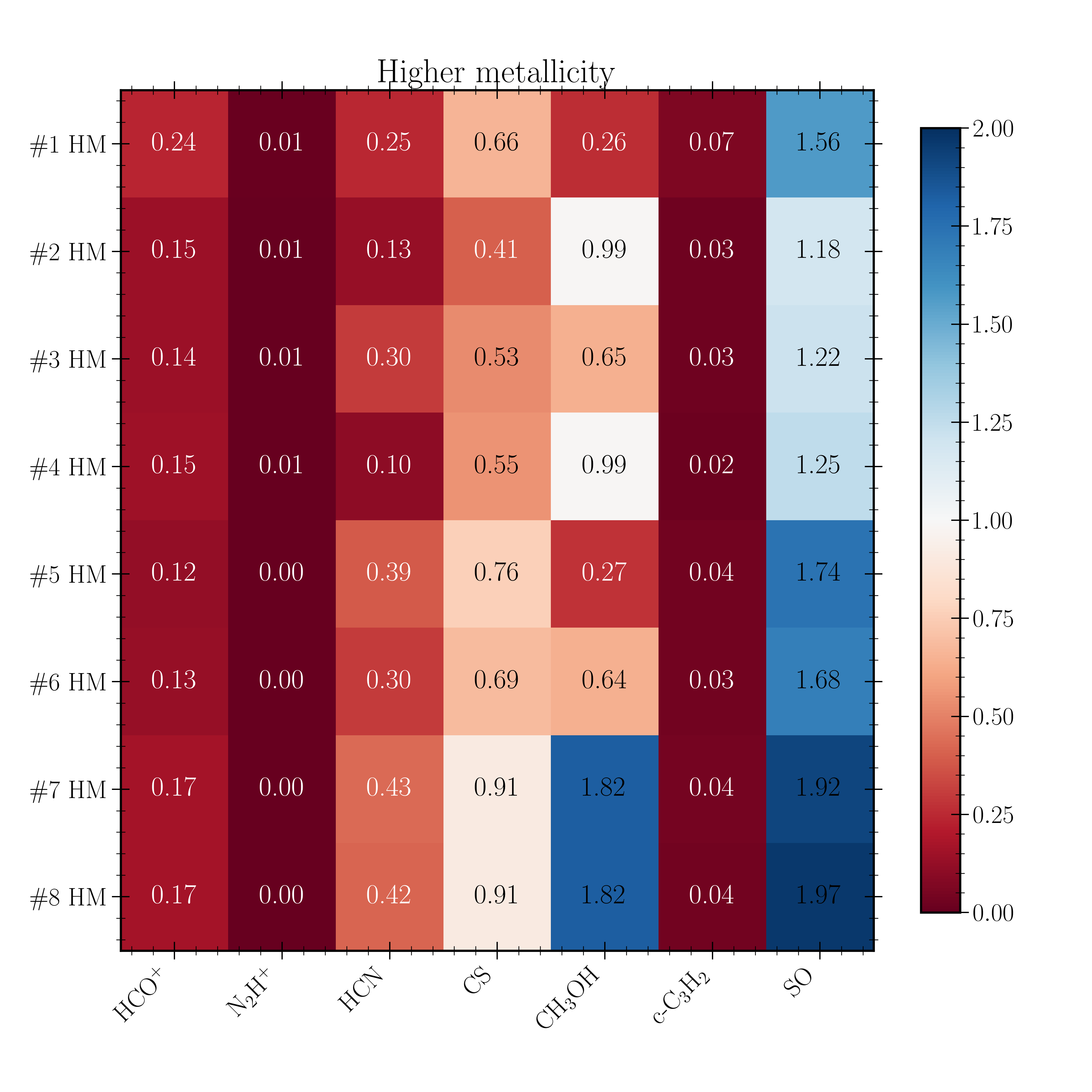}}
  \caption{Similar to Fig. \ref{fig:ratio_xy_SM} but for higher metallicities.}
     \label{fig:ratio_xy_HM}
\end{figure}

\begin{figure}[ht]
\resizebox{\hsize}{!}
        {\includegraphics{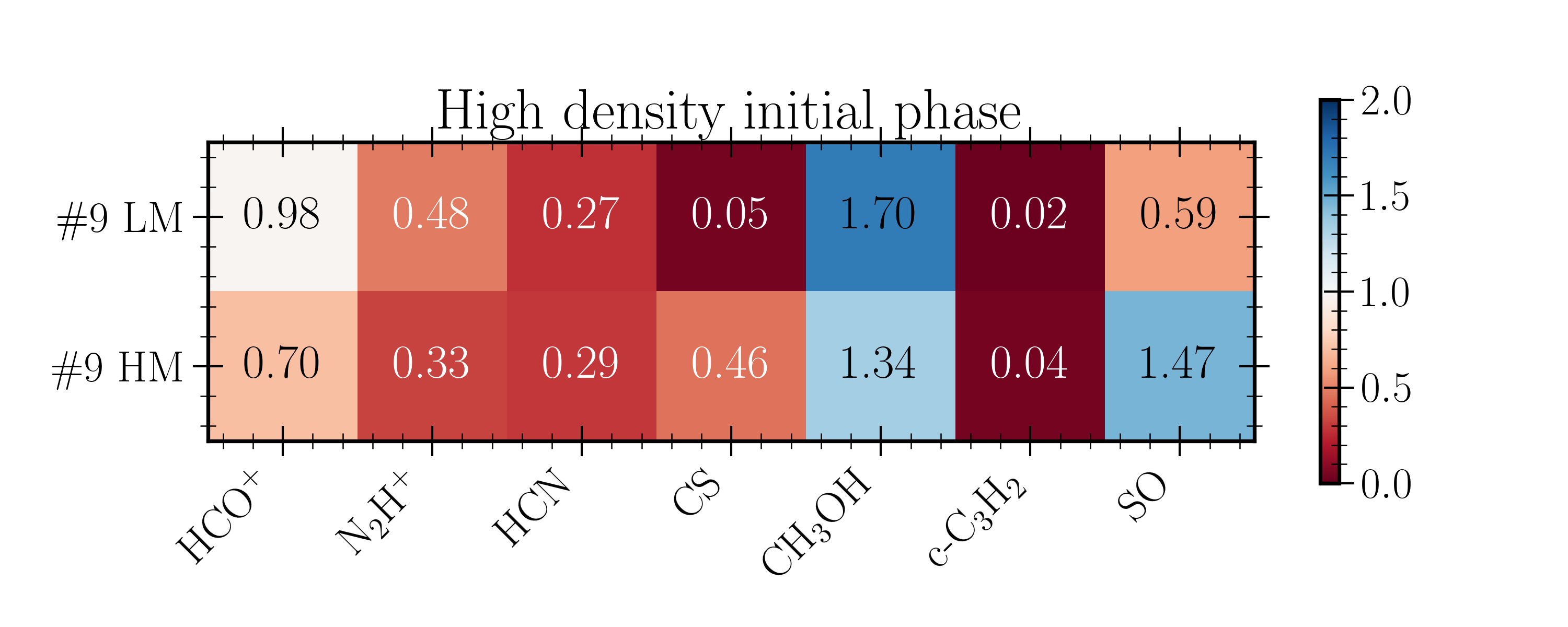}}
  \caption{Comparison table for the peak intensity of the modeled emission lines toward L1544 (dust peak position) and toward the dust peak in the $x$-$y$ plane. This figure shows the comparison for the model \#9 with either standard or higher metallicities.}
     \label{fig:ratio_xy_1e4}
\end{figure}

Figures \ref{fig:ratio_xy_SM} and \ref{fig:ratio_xy_HM} show the ratio of the peak emission observed toward the continuum emission peak of L1544 and the values extracted from the model for models \#1-\#8 with LM and HM abundances. 
Several trends are notable. The intensity of the HCO$^{+}$ 1--0 transitions are well reproduced by models with standard metallicities. For the HM models, on the other hand, the intensities are too weak by factors of 4--10. The intensity of the central component of the N$_2$H$^{+}$ 1--0 hyperfine transition is weaker by factors of $\geq$10 for the LM and HM models. 
The intensity of the central component of the HCN 1--0 hyperfine transition is underproduced by factors of 2-10 in the LM and HM models. Sulfur-bearing species are underproduced in the LM models, but perform reasonably well in the HM models, which can be directly understood as a consequence of the higher initial abundance of sulfur \citep{2019A&A...624A.108L}. 
For methanol, the differences between the two metallicities are minimal, which is due to the similar abundances of carbon and oxygen in the LM and HM models. However, the carbon-chain molecule $c$-C$_3$H$_2$ is better reproduced in models \#6-\#8 by the HM models. Nonetheless, $c$-C$_3$H$_2$ remains underproduced in all models. 
Models \#1-\#8 all agree poorly with the observed intensities. To assess whether this is linked to a too diffuse initial phase, we also ran an additional set of models with a denser initial phase and higher shielding (\#9) to determine whether this would lead to better agreement. The results are shown in Fig. \ref{fig:ratio_xy_1e4}. Overall, a denser initial phase does lead to better agreement, but N$_2$H$^{+}$ and $c$-C$_3$H$_2$ remain weaker than expected. HCO$^{+}$ and the sulfur-bearing species also differ. The former is best reproduced with LM abundances, while SO and CS are significantly closer to the observed intensities with the HM conditions.

\begin{figure}[ht]
\resizebox{\hsize}{!}
        {\includegraphics{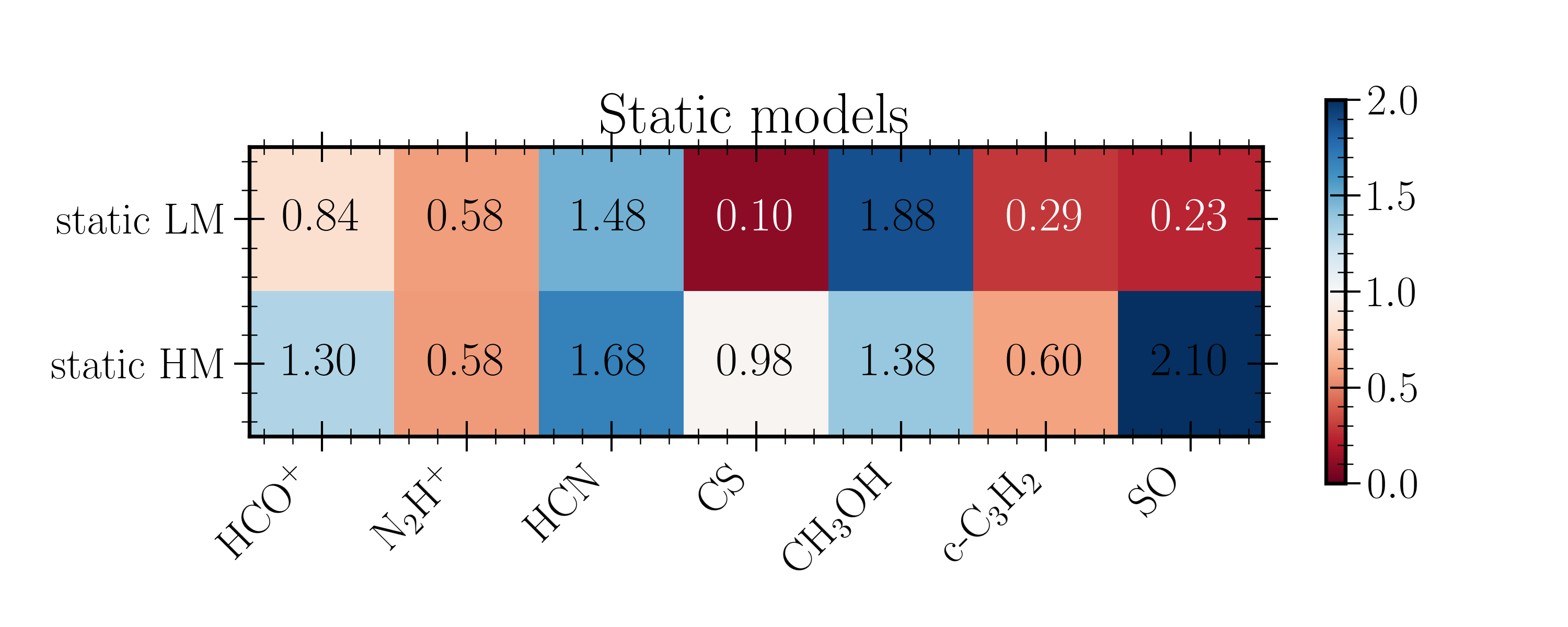}}
  \caption{Comparison table for the peak intensity of the modeled emission lines toward L1544 (dust peak position) and toward the dust peak in the $x$-$y$ plane. This figure shows the comparison for the static model with either standard or higher metallicities.}
     \label{fig:ratio_xy_static}
\end{figure}

Figure \ref{fig:ratio_xy_static} shows the comparison of the peak intensities in the fiducial static model for both LM and HM models. In this case, a better agreement with the observed ratios toward L1544 is observed for all species. In particular, the higher-metal model is within a factor of $\sim$~2 in intensity for all molecules.

Several of the targeted transitions are optically thick, with $\tau > 0.5$. The comparable peak intensities therefore do not directly relate to the total column density, but instead indicate that the combination of the density, abundance, and temperature profile of the modeled core leads to excitation conditions that are qualitatively similar to L1544.

\section{Discussion}  \label{sec:discussion}
\subsection{Effects of dynamic versus static modeling on intensity predictions}

One of the principal motivations for us to include the dynamic tracer particle trajectories instead of  relying on a static approach was to assess whether this time-dependent treatment leads to an improved agreement with observational constraints, particularly in terms of the spatial morphology of the emission regions and the relative intensities of key molecular species.

Our results do not lend themselves to a straightforward conclusion. 
With the updated chemical network, the predicted emission morphology of CH$_3$OH and $c$-C$_3$H$_2$ is qualitatively similar for the dynamical model to the observed distribution toward L1544, whereas the predicted morphology for the static model is notably different. 
This is different for the predicted intensities. The dynamical models (\#1--\#8) predict lower intensities for all molecules when compared with the static model for similar initial elemental abundances. The lower abundances and intensities predicted by the dynamical model can be attributed to a central limitation in the current setup: the dynamical tracer particles follow trajectories over a total duration of 300~kyr, resulting in shorter residence times at high densities than in the static model, where chemical evolution is computed over 1~Myr across the entire parameter space. While a shorter residence time agrees better with our current understanding of a dynamic star formation process, it explains why the static model produces more N$_2$H$^+$, which requires substantial time to form at higher densities \citep{2021A&A...646A..97T, 2023MNRAS.524.5971P}. For dynamical models with a longer and/or a denser initial phase (\#4 and \#9), the N$_2$H$^+$ emission is brighter, although still not as bright as in the static model. An even denser initial phase is likely to bring the abundances and intensities of the dynamical model closer to that of the static model, but this comes with additional caveats.

In the current setup, this challenge cannot be easily circumvented. The core we studied here reaches central densities comparable to those of L1544 after 300~kyr of evolution and evolves to form a protostar shortly thereafter. Adopting a denser initial phase improves the agreement with observations, but at the cost of compromising the self-consistency of the model. The temperature profile of L1544 is well reproduced using an ambient extinction of $A_\mathrm{v,eff}^\mathrm{amb} \simeq2$~mag, which is consistent with a lower background density of $n_\mathrm{H}\lessapprox$10$^{3}$~cm$^{-3}$. Increasing $A_\mathrm{v,eff}^\mathrm{amb}$ to $\simeq4$~mag, consistent with densities  $n_\mathrm{H}\gtrsim$10$^{4}$~cm$^{-3}$, leads to temperatures that are too low at the edge of the core (see Fig. \ref{fig:temp_profiles}), which contradicts the previous models of L1544 and the temperature maps derived from \emph{Herschel}/SPIRE continuum maps \citep{2010MNRAS.402.1625K, 2023ApJ...944..208B}. 
Given this inconsistency, we are cautious to suggest a denser ambient medium with higher shielding as a realistic option. An alternative approach might be to extend the duration $t_\mathrm{init}$  of the initial static phase in the dynamical models beyond 2~Myr. However, increasing $t_\mathrm{init}$ from 1~Myr to 2~Myr between models \#3-\#4 leads to a rather modest increase in the intensity of N$_2$H$^+$ by 25\%. This suggests that increasing the duration alone does not lead to significantly higher abundances of N$_2$H$^+$; higher densities are necessary as well.

In contrast, the static model predicts stronger intensities for all molecules. The predicted intensities are notably closer to what is observed toward L1544, except for CH$_3$OH, which is generally reproduced well by both approaches. The HM static model predicts intensities that are all within a factor of $\sim$2 from the observed intensities (Fig. \ref{fig:ratio_xy_static}). This level of agreement is quite satisfactory given that the modeled core is not a perfect replica of L1544. This suggests that, despite their simplified assumptions about physical evolution, static models can still serve as reliable tools for studying pre-stellar cores in three dimensions, especially when the primary goal is to reproduce observed emission intensities. 

For $c$-C$_3$H$_2$, the intensity is underproduced by a factor of $>$10 in the dynamical models. For the static models, the agreement is within a factor of $\sim$4, but the intensity and abundance are still underproduced. This behavior is a consequence of a rather low steady-state abundance of $c$-C$_3$H$_2$ in the chemical network. At higher densities and longer chemical timescales, carbon and hydrogen are locked in COMs and the abundance of $c$-C$_3$H$_2$ is reduced. The effect is less pronounced in the static model because the outer less dense regions of the core remain chemically young, meaning that a smaller fraction of the carbon is locked in CO. In the dynamical models, all trajectories experience periods of higher densities ($n_\mathrm{H}\gtrsim $10$^{4}$~cm$^{-3}$), which leads to a lower $c$-C$_3$H$_2$ abundance, and the effect is not reversed when the particles enter more diffuse regions later on.

For N$_2$H$^{+}$ and $c$-C$_3$H$_2$, the reaction rates used here from KIDA2014 were recently updated in KIDA2024 \citep{2024A&A...689A..63W}. The update contains revised rates that contribute to the formation and destruction of both these species and may consequently affect the results presented here. Updating the chemical network used here to the latest rates and benchmarking the updates is the subject of future work. 

Future studies could focus on another of the ~200 protostars that form in the simulation. However, when longer tracer particle tracks are used, another limitation might play a role. As the 4~pc$^{3}$ RAMSES domain has periodic boundary conditions, the tracer particles will pass the domain multiple times when longer simulation timescales are considered. This may not pose a problem compared to very dense star-forming clusters, but it may not represent a realistic scenario if the simulation is compared to star-forming clouds with lower densities. 

We caution that these results do not necessarily reflect that dynamical models in general are worse than static models, but rather that the current setup presented here lacks a long enough dynamical timescale to reproduce the physico-chemical history of the gas well enough to produce more realistic results than the static model. \citet{2022A&A...668A.131S} studied the effects of dynamical and static 1D models for L1544 and found generally similar results for the static and dynamical models, although the dynamic models did produce a slightly better agreement with observed line intensities.

We did not include a comparison of the line profiles between the model and L1544. The specific shape of the emission line profiles computed for the 3D model are highly dependent on the viewing angle, although the peak intensities remain comparable. A deeper analysis of the effect of the viewing angle on the modeled spectra is left for future work.

\subsection{Interpretation of the morphologies}
While the static model reproduces the intensities toward L1544 well, it fails to capture the observed characteristic structure for CH$_3$OH and $c$-C$_3$H$_2$ toward L1544 and other young cores. In contrast, the dynamical models presented here reproduce the observed dichotomy in the molecular emission maps better. Because both approaches employ the same underlying physical structure and chemical network, the differences in morphology are attributed to the distinct evolutionary histories encoded in the tracer particle trajectories as opposed to the chemical evolution of the static model at fixed physical conditions. The insensitivity of the modeled morphology to the initial chemical conditions further supports the suggestion that the spatial distributions are primarily governed by the trajectories of individual tracers and not by the initial chemical state. 
Correlation analyses of column densities and integrated physical parameters reveal that CH$_3$OH and $c$-C$_3$H$_2$ correlate with higher visual extinction and H$_2$ column densities. This correlation is stronger for methanol in models with shorter initial chemical phases, suggesting that the effect of individual tracer trajectories diminishes when the initial chemistry is allowed to evolve over longer timescales. 
The correlation for methanol is stronger in models with lower ambient shielding, indicating that the observed offset in methanol emission (particularly its concentration in more shielded regions) is strongly affected by radiative processes. Higher ambient extinction leads to lower temperatures and diminished differences in photochemical processing along different particle paths. These conditions promote efficient CO ice formation and subsequent hydrogenation, enhancing methanol production, which contributes to the offset seen in the maps (Fig.  \ref{fig:chem2plot_fidHM}). This finding agrees with previous studies by \citet{2016A&A...592L..11S,2020A&A...643A..60S}, who concluded that methanol preferentially traces the most shielded parts of the core. It is worth noting that the correlation between the local physical parameters ($N$(H$_2$) and $\hat{N}(A_\mathrm{v,eff})$) and the molecular emission does not indicate a reduction of the dependence on the initial conditions and tracer particle history. Instead, the observed correlation suggests that gas in regions with a higher density and visual extinction have, as part of their tracer particle histories, undergone a physical evolution promoting formation of, for example, CH$_3$OH. Conversely, regions with lower density and visual extinction in the core studied here tend to have physical histories that are less suitable for CH$_3$OH formation. This is different in the static model, where the abundance of CH$_3$OH in any given point of the model only depends on the local physical conditions at that specific point.
Compared with the intensity of $c$-C$_3$H$_2$ observed toward L1544, the static and dynamic models are too faint, which suggests that some aspects of both models fail to capture the formation of this molecule.

\subsection{Comparison of the metallicities of sulfur-bearing species}
Throughout this work, we have compared two sets of initial abundances. The standard metallicities presented here are commonly used in a broad range of astrochemical studies of pre-stellar cores \citep{2001ApJ...552..639A, 2017ApJ...842...33V, 2021A&A...656A.109R}. They are depleted in sulfur compared to the solar values \citep{1993ApJ...413L..51F, 2009AJ....138.1577D}. In our simulations, the modeled intensities of CS and SO are comparable (within factors of $\lesssim$2--3) to observed values for L1544 using the un-depleted (higher) metallicities, while the depleted abundances underproduce the intensities by a factor of $\gtrsim$5. A similar result was also reported in \citet{2024MNRAS.531.4408P} for a 3D smoothed-particle hydrodynamic simulation of a star-forming cloud and by \citet{2019A&A...624A.108L}, who studied the depletion of sulfur from diffuse to dense clouds. Likewise, \citet{2022A&A...658A.168H} reported little evidence of sulfur depletion across a sample of young cores, with L1544 showing slightly more depletion than a number of younger cores.
These results suggest that the high sulfur depletion (a factor of $\sim$200) introduced in the standard metallicities may be a poor choice for modeling starless and pre-stellar cores. The initial motivation behind the depleted sulfur abundance was a series of studies that showed a clear deficit in sulfur-bearing molecules compared to cosmic abundances \citep{2015MNRAS.450.1256W}. This led to the suggestion that sulfur might be locked in a unidentified solid phase \citep[e.g.,][]{2012ApJ...751L..40J, 2016A&A...585A.112M, 2025ApJ...986L..17F}. However, based on current astrochemical models, a high sulfur depletion appears unnecessary in L1544 and several other cores. This agrees with the recent study by \citet{2023A&A...670A.114F}, who compared observations and astrochemical models of CS across various star-forming regions. They found that the extent of sulfur depletion likely varies between star-forming clouds due to environmental effects. The regions studied in that work required a more modest depletion, with depletion factors of $\sim$10, rather than $\sim$100 as in the LM models presented here.

\subsection{Caveats of the model}

As discussed in earlier sections, the underproduction of the species in question within the dynamical model is likely due to the specific initial chemical conditions, such as the density, temperature, and the duration of the initial phase. Although the current model incorporates a more realistic representation of core formation by accounting for the physical evolution of the gas within the core, it remains sensitive to the initially selected conditions. This dependence on initial conditions is a common challenge in astrochemical modeling \citep[e.g.,][]{2019MNRAS.486.5197V, 2020ApJ...897..110A}. Overcoming this limitation would necessitate longer 3D cloud simulations, capable of fully capturing the transition from the diffuse interstellar medium through molecular cloud formation to the development of dense cores.

In the present model, we assumed a perfect coupling between gas and dust temperatures. This assumption is not valid at the lower densities present in the outer parts of the core and might specifically affect the conditions in the lower-density regions of the core, where carbon-chain molecules such as $c$-C$_3$H$_2$ are more abundant. However, this assumption applies to the static and dynamical models and therefore does not explain the poorer agreement for the modeled intensities in the dynamical model.

One possible problem of the tracer particle approach we used is that the tracer particles may not adequately  sample all regions of parameter space. The current setup includes $\sim$20,000 tracer particles. These predominantly trace the denser regions of the gas, which can contribute to the underproduction of $c$-C$_3$H$_2$ in the model as $c$-C$_3$H$_2$ is destroyed in higher-density regions. It is computationally challenging to circumvent this behavior. The underlying RAMSES model already includes a high number of tracer particles ($>10^{9}$), but not all particles will trace any given core in the molecular cloud simulation. To assess to which extent insufficient sampling may affect the results, we compared the densities traced by the tracer particles with the densities of the underlying grid. They agree well overall, but reveal slight differences at lower densities that might contribute to a poorer sampling of the chemistry in these regions of the dynamical models (see Figs. \ref{fig:interp_compare} and \ref{fig:interp_compare2}).

\section{Summary}  \label{sec:5}
We presented a dynamical 3D physico-chemical model of a pre-stellar core resembling the pre-stellar core L1544. We compared the chemical morphology and modeled molecular intensities with L1544 and a static physico-chemical model of the same core. 
Our results are listed below.
\begin{itemize}
        \item Qualitatively, the dynamical model reproduced the chemical structure of carbon chains and COMs (traced by $c$-C$_3$H$_2$ and CH$_3$OH, respectively) as observed in L1544. This was not the case for the static model. This effect was stronger when a weaker ambient extinction was applied around the core, indicating that it is directly linked to the structure of the core as traced by individual tracer particles.
        \item The dynamical model was unable to reproduce the observed intensities of the modeled molecules simultaneously. Instead, the model had to be fine-tuned for individual molecules to produce a good agreement with observations of L1544. This was not the case for the static model, where the modeled species all agreed simultaneously within a factor of $\sim$2.
        \item The weak molecular intensities predicted by the dynamical model are primarily due to the relatively short evolutionary timescale of the tracer particle trajectories. In particular, the limited duration prevents a sufficient formation of N$_2$H$^{+}$ (a key tracer of high-density regions), leading to underproduction compared to observations. One potential remedy is to pre-evolve the initial chemical composition of the tracer particles at higher densities, which helps to reduce the disagreement with the observed L1544 data and the static model. However, this approach is inconsistent with the inferred temperature profile of L1544, thereby compromising the physical self-consistency of the model.
        \item The intensities derived from the dynamical model are highly sensitive to the conditions and duration of the initial static phase, reflecting a high degree of chemical inheritance. While this underscores the importance of the initial state in shaping the final chemical composition, it also highlights the need for careful and physically motivated choices when setting the initial conditions in astrochemical modeling.
        \item Models with cosmic sulfur abundance agree well with observed intensities of SO and CS, while depleted abundances lead to underestimated intensities of the two molecules.

\end{itemize}

The static model we presented agrees well with the observed intensities toward L1544. However, the intensities predicted by the dynamical model using tracer particles are too low for most molecules, suggesting that some critical aspects are not adequately accounted for in the current framework. Future work should explore alternative approaches to improve the agreement for all molecules in the dynamical model. This might involve adjusting the initial phase or studying longer trajectories of the tracer particles to better capture the chemical evolution.

\begin{acknowledgements}
We wish to thank the anonymous referee for helpful comments which improved the manuscript.
S.S.J. and S.S. wish to thank the Max Planck Society for the Max Planck Research Group funding. All other authors affiliated to the MPE wish to thank the Max Planck Society for financial support.
The research leading to these results has received funding from the Independent Research Fund Denmark through grant No. DFF 10.46540/4283-00305B (TH). The Tycho supercomputer hosted at the SCIENCE HPC center at the University of Copenhagen, was used for carrying out the simulations, the analysis, and long-term storage of the results. This research has made use of spectroscopic and collisional data from the EMAA database (https://emaa.osug.fr and https://dx.doi.org/10.17178/EMAA). EMAA is supported by the Observatoire des Sciences de l'Univers de Grenoble (OSUG).  This paper makes use of {\sc matplotlib} \citep{Hunter:2007} and {\sc scipy} \citep{2020SciPy-NMeth}.
\end{acknowledgements}

\bibliographystyle{aa}
\bibliography{b.bib}

\begin{appendix}

\section{The physical structure of the core}\label{app:3D_core}
\begin{figure*}[ht]
\resizebox{\hsize}{!}
        {\includegraphics{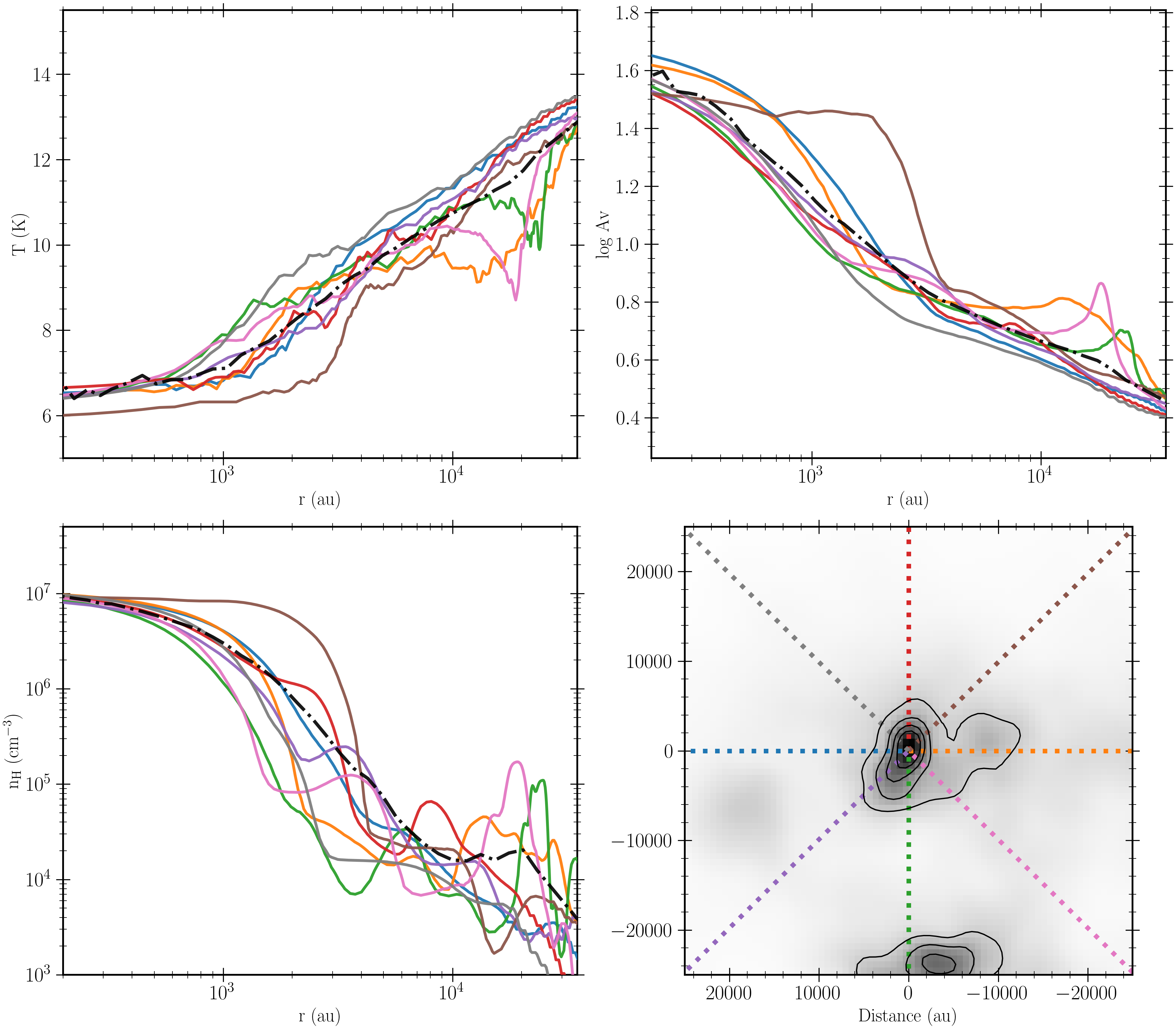}}
  \caption{Physical structure of the core in the $x$-$y$ plane. \emph{Left and top right}: Radial profiles for the dust temperature, number density and visual extinction are shown along 8 different radial cuts on the x-y plane.
  \emph{Lower right}: Continuum image of the core at 1.1~mm (in grayscale) and 30\%, 45\%, 60\%, and 75\% of the peak intensity (contours). Colored lines indicate the direction along which the radial profiles in the remaining panels were extracted.}
     \label{fig:3D_core}
\end{figure*}

Figure \ref{fig:3D_core} shows radial profiles along 8 directions for the simulated core marked with different colors. The map in the lower right corner shows the continuum emission at 1.1~mm and the 8 directions in which the profiles were extracted. In this figure, an external extinction of $(A_\mathrm{v,amb})$ = 2~mag was adopted.

\section{Example of tracer particle trajectories}\label{app:trajectories}
\begin{figure*}[ht]
\resizebox{\hsize}{!}
        {\includegraphics{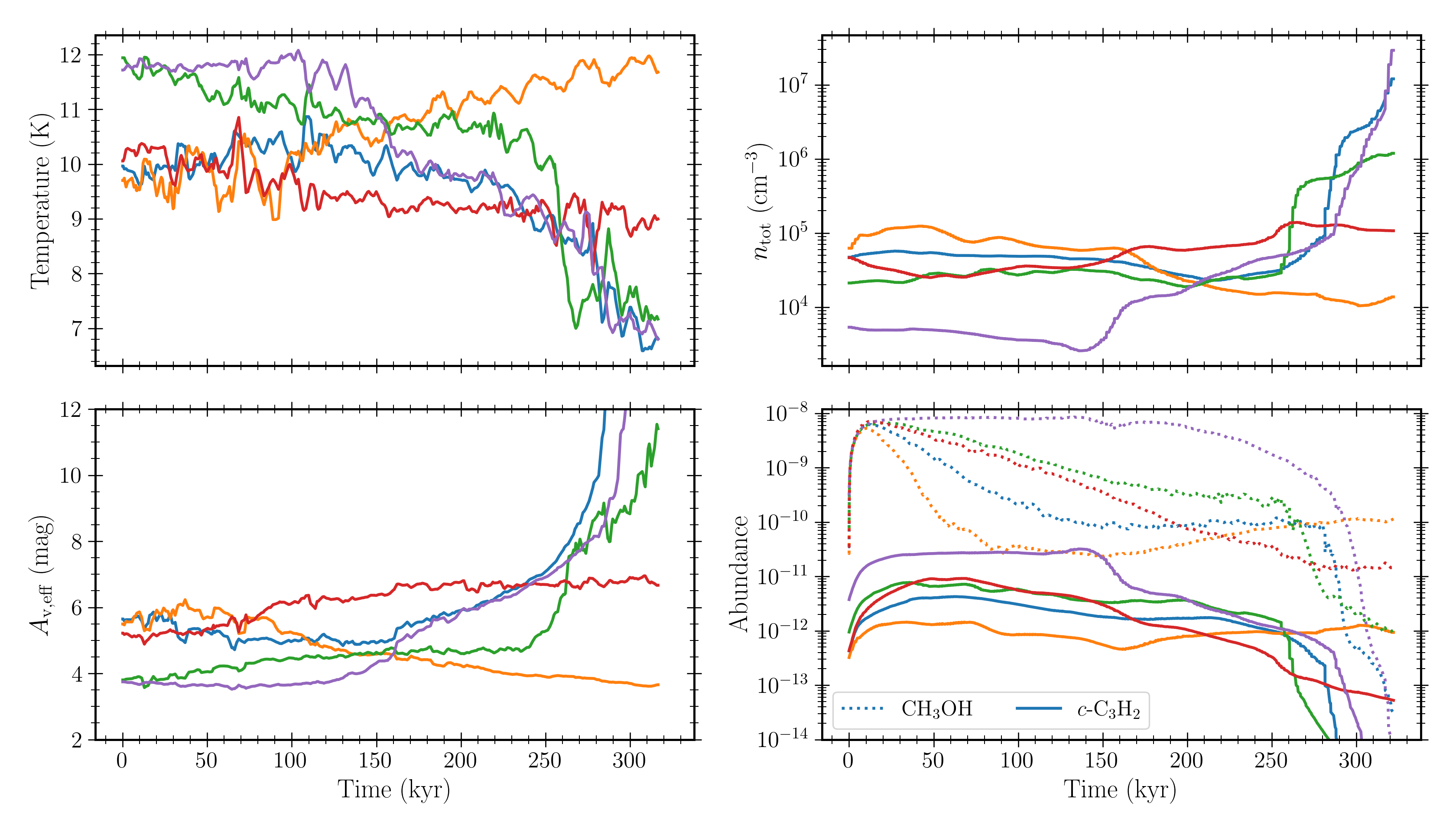}}
  \caption{Trajectories for five tracer particles. Abundances are shown for model \#3 LM.}
     \label{fig:tracer_trajectories}
\end{figure*}

In Fig. \ref{fig:tracer_trajectories} we show an example of five tracer particle trajectories. The corresponding gas-phase abundance of CH$_3$OH and $c$--C$_3$H$_2$ are shown in the lower-right panel. The tracer particle tracjectories start at different position, with different physical conditions. All tracer particles except one (orange color) finish at higher densities than they started, with varying degree of freezeout of both gas-phase species.

\section{Testing the sampling of the tracer particles}\label{app:sampling}
\begin{figure}[ht]
\resizebox{\hsize}{!}
        {\includegraphics{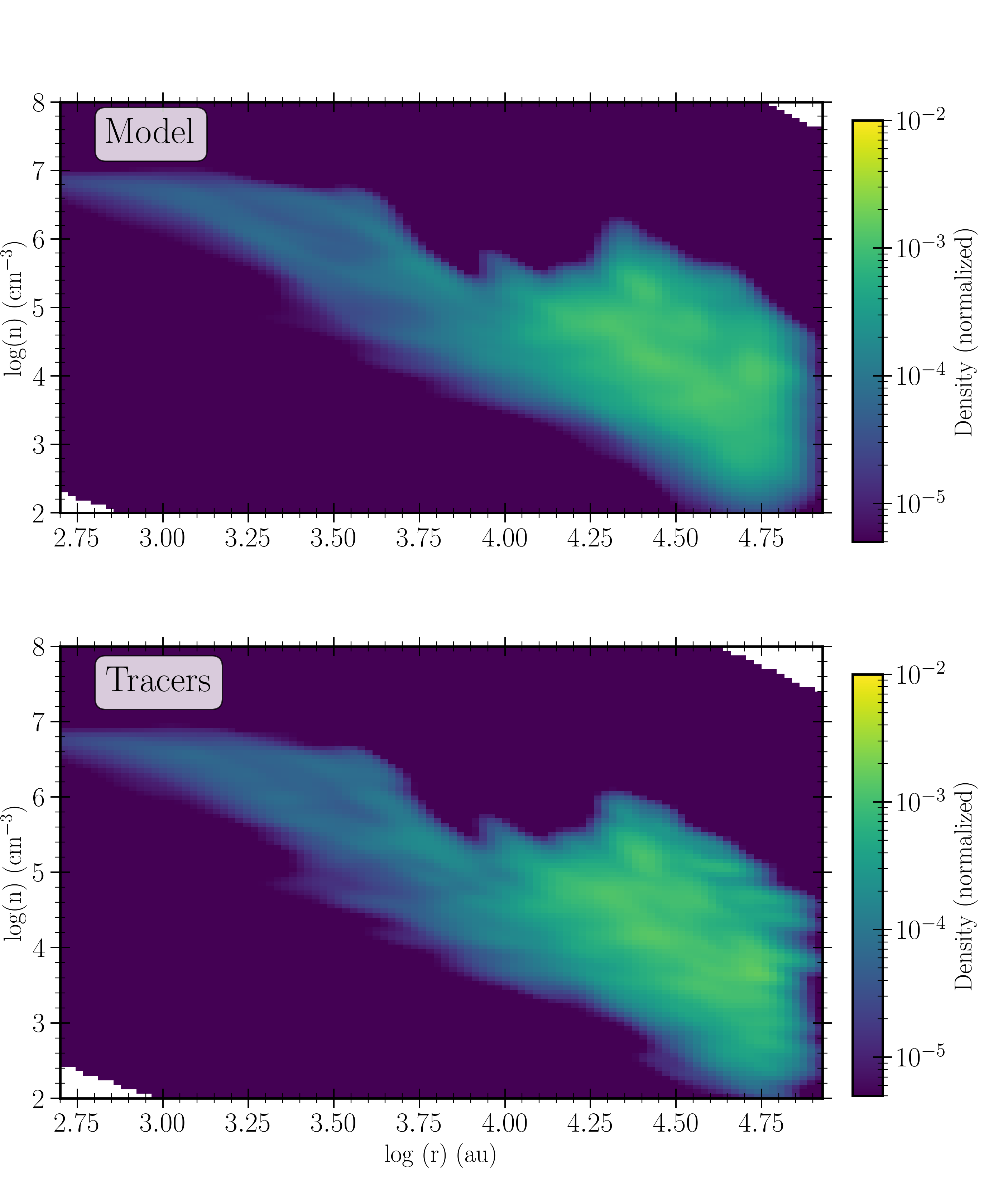}}
  \caption{\emph{Top:} KDE for the density structure of the core as calculated using the true grid. \emph{Bottom:} KDE for the density structure of the core when interpolated from the density of the tracer particles.}
     \label{fig:interp_compare}
\end{figure}

\begin{figure}[ht]
\resizebox{\hsize}{!}
        {\includegraphics{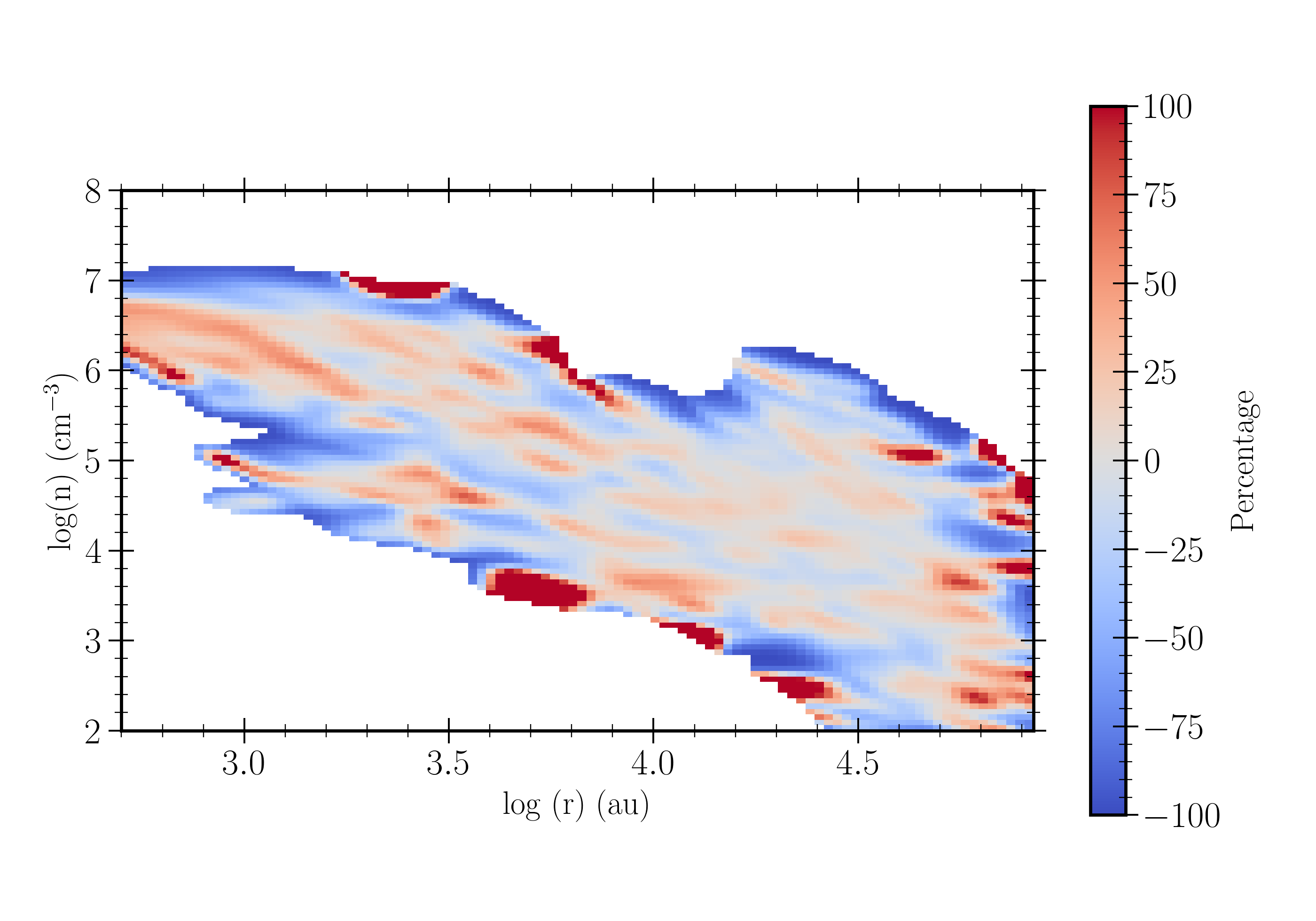}}
  \caption{Deviation between the density structure of computational RAMSES grid and the density structure when interpolated from the $\sim$20,000 tracer particles used in this study. The figure shows the deviation between the kernel density estimates for the density structure derived from either grid or tracer particles. Positive value indicate that the density derived from the tracer particles exceed the local grid value.}
     \label{fig:interp_compare2}
\end{figure}

To visualize how well the tracer particles sample the actual core, we compared the structure of the 3D core when interpolated from tracer particle positions with the actual computational grid.
Figure \ref{fig:interp_compare} visualizes a kernel density estimation (KDE) of the density structure based on the actual model (top panel) and the KDE derived after interpolating the tracer particle densities onto the grid (bottom panel). The figures shows minor deviations along the edges of the KDE, while the overall density distributions are almost identical. Figure \ref{fig:interp_compare2} shows the deviation between the KDE and the underlying grid in percent.

\section{References for spectroscopic data and collisional rates for radiative transfer calculations}\label{app:1}
An overview of the transitions studied using the nonlocal thermodynamic equilibrium line radiative transfer modeling is included in Table \ref{tab:lines}. The table also list references for the collisional rates used for the modeling.

\begin{table*}
\caption{Transitions used for the line radiative transfer modeling.}
\label{tab:lines}
\centering
\begin{tabular}{lccc}
\hline 
Species & Frequency (GHz) & Transition & References \\ \hline
HCO$^{+}$ &     89.1885          &  $J$=1--0 &  \citet{2007ApJ...669L.113T, 2020MNRAS.497.4276D} \\
N$_2$H$^{+}$ &      93.1735          & $J$=1--0 & \citet{2012ApJS..203...11C, 2015MNRAS.446.1245L} \\
CS      &      97.9810          & $J$=2--1 & \citet{2005JMoSt.742..215M, 2018MNRAS.478.1811D} \\      
SO      &      99.2999           & $N$,$J$ = (2,3)--(1,2) & \citet{1976JMoSp..60..332C, 2007JChPh.126p4312L}\\
CH$_3$OH       &      96.7394           &  $J_{Ka,Kc}$ = 2$_{1,2}$-1$_{1,1}$ ($E2$) & \citet{2016JMoSp.327...95E, 2010MNRAS.406...95R}\\
$c$--C$_3$H$_2$ &      84.7277           &  $J_{Ka,Kc}$ = 3$_{2,2}$-3$_{1,3}$ & \citet{1998JQSRT..60..883P, aaa} \\
HCN       &         88.6318      &  1--0  &  \citet{2002ZNatA..57..669A, 2017MNRAS.468.1084H}  \\
C$^{18}$O & 219.5604, 329.3306  & 2--1, 3--2 & \citet{2010ApJ...718.1062Y} \\
\end{tabular}
\end{table*}

\section{Temperature profiles as a function of ISRF extinction}\label{app:T_profiles}
\begin{figure}[ht]
\resizebox{\hsize}{!}
        {\includegraphics{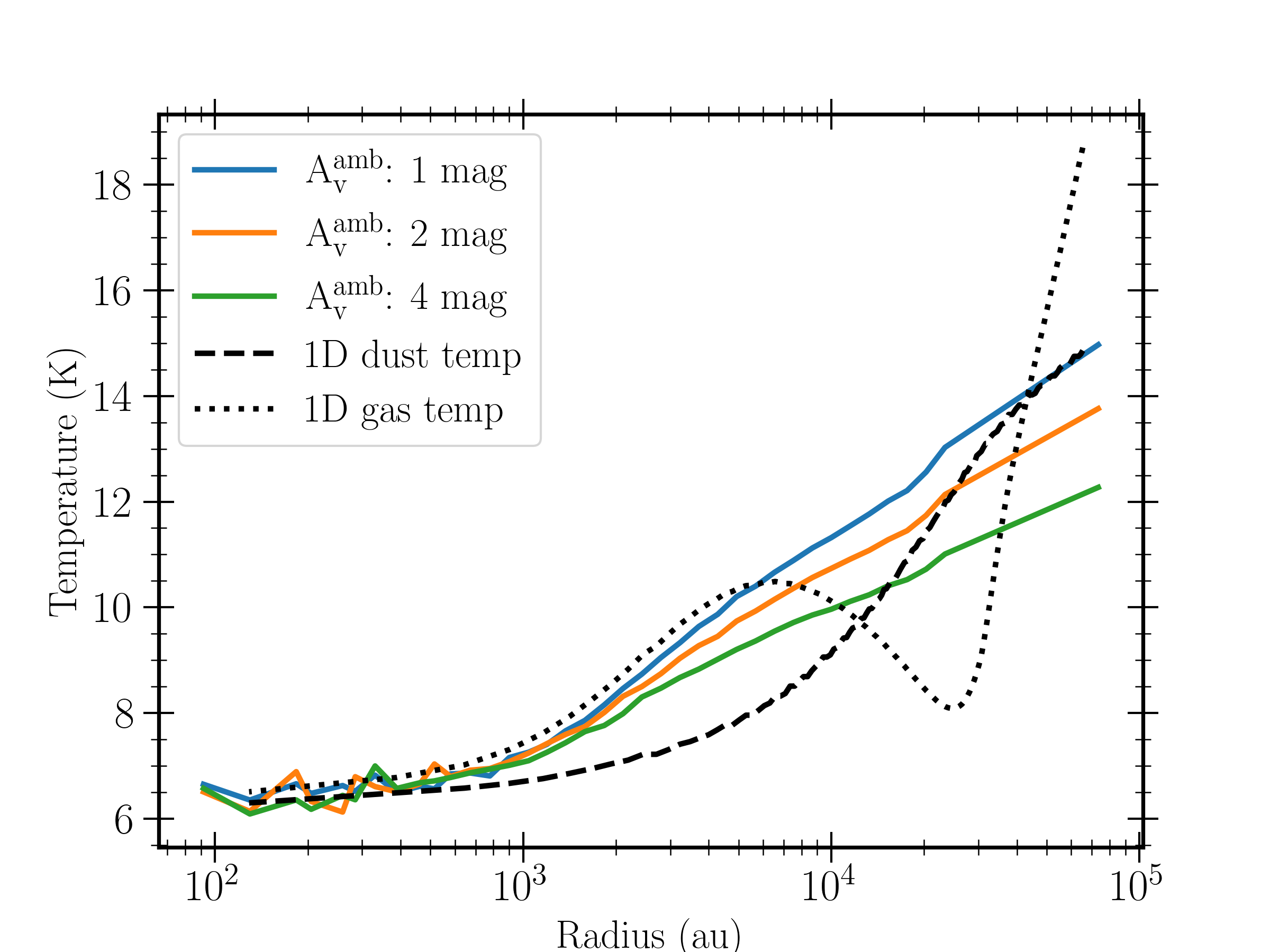}}
  \caption{Mean temperature profiles for different degrees of ISRF attenuation. Black lines show the gas and dust temperature profiles from the 1D L1544 presented in \citet{2015MNRAS.446.3731K}.}
     \label{fig:temp_profiles}
\end{figure}

Figure \ref{fig:temp_profiles} shows a comparison of the dust temperatures computed for different external shielding: $A_\mathrm{v,eff}^\mathrm{amb}$=~1~mag, $A_\mathrm{v,eff}^\mathrm{amb}$=~2~mag, and $A_\mathrm{v,eff}^\mathrm{amb}$=~4~mag. Also included are the gas and dust temperature profiles from the 1D model of \citet{2015MNRAS.446.3731K}.

\end{appendix}
\end{document}